\documentclass[journal, draftcls, draftclsnofoot, onecolumn,12pt]{IEEEtran}

\usepackage{amsmath, amsthm, amsfonts, amssymb, amsbsy}
\usepackage{bm} %boldface in math mode
\usepackage{setspace}
\usepackage{graphicx}
\usepackage{algorithm}
\usepackage{multirow}
\usepackage{subfigure}
\usepackage{verbatim}
\usepackage{graphicx,pstricks,pspicture,cite}

\newcommand{\twotriangle}{\hfill $\bigtriangleup \bigtriangleup$  }
\newcommand{\eax}{\twotriangle  \end{example}}
\newcommand\bim{\begin{itemize}}
\newcommand\eim{\end{itemize}}

\newcommand\bH{{\bf H}}
\newcommand\bM{{\bf M}}
\newcommand\bN{{\bf N}}

\newcommand\bA{{\bf A}}

\newcommand\bE{{\bf E}}
\newcommand\bI{{\bf I}}
\newcommand\bG{{\bf G}}
\newcommand\bS{{\bf S}}

\newcommand\bZ{{\bf Z}}
\newcommand\bD{{\bf D}}
\newcommand\bB{{\bf B}}
\newcommand\bP{{\bf P}}
\newcommand\bQ{{\bf Q}}
\newcommand\bV{{\bf V}}

\newcommand\bJ{{\bf J}}
\newcommand\bX{{\bf X}}

\newcommand\bh{{\bf h}}

\newcommand\bx{{\bf x}}

\newcommand\bz{{\bf z}}

\newcommand{\cG}{\mathcal{G}}

\newcommand{\cGD}{\cG [\Delta]}

\newtheorem{thm}{Theorem}
\newtheorem{exmp}{Example}
\newtheorem{defn}{Definition}

\begin{document}
\title{New Classes of Partial Geometries and Their Associated LDPC Codes}

%\vspace{-30pt}

\author{Qiuju Diao, \IEEEmembership{Member, IEEE}, Juane Li, \IEEEmembership{Student Member, IEEE}, \\
Shu Lin, \IEEEmembership{Life Fellow, IEEE}, and Ian Blake, \IEEEmembership{Life Fellow, IEEE}\\

\thanks{Q. Diao is with Sandisk Corp., Milpitas, CA 95035}
\thanks{Email:judiao@ucdavis.edu}
\thanks{J. Li and S. Lin are with ECE Dept., UC Davis, Davis, CA 95616}
\thanks{Email: \{jueli, shulin\}@ece.ucdavis.edu}
%\thanks{S. Lin is with ECE Dept., UC Davis, Davis, CA 95616-5294}
%\thanks{Email: shulin@ece.ucdavis.edu}
\thanks{Ian F. Blake is with ECE Dept. UBC, Vancouver, BC V6T 1Z4}
\thanks{Email: ifblake@ece.ubc.ca}
\thanks{This work was supported by the NSF under Grant CCF-1015548.}
}
\maketitle

\begin{abstract}
The use of partial geometries to construct parity-check matrices for LDPC codes
has resulted in the design of successful codes with a probability
of error close to the Shannon capacity at bit error rates down to $10^{-15}$.
Such considerations have motivated this further investigation. A new and simple construction of a type of partial geometries with quasi-cyclic structure is given and their properties are investigated. The trapping sets of the partial geometry codes were considered previously using the geometric aspects of the underlying structure to derive information on the size of allowable trapping sets. This topic is further considered here. Finally, there is a natural relationship between partial geometries and strongly regular graphs. The eigenvalues of the adjacency matrices of such graphs are well known and it is of interest to determine if any of the Tanner graphs derived from the partial geometries are good expanders for certain parameter sets, since it can be argued that codes with good geometric and expansion properties might perform well under message-passing decoding.
\end{abstract}
 
\begin{IEEEkeywords}
LDPC codes, partial geometries, strongly regular graphs, protographs, expander graphs
\end{IEEEkeywords}
%\newpage

\section{Introduction} \label{sect1}

Partial geometries play an important role in the construction of low-density 
parity-check (LDPC) codes \cite{gall,mac,mac1} which currently give the most promising 
coding technique for error control in communication and data storage systems 
due to their capacity-approaching performances and practically implementable 
decoding algorithms. Partial geometries are members of a broad class
of combinatorial configurations with geometric properties referred to as finite 
geometries. The first classes of LDPC codes based on partial geometries were constructed 
based on Euclidean and Projective geometries over finite fields \cite{kou}.  
These classes of finite geometry LDPC codes have an abundance of algebraic 
and geometric structures and perform well with iterative decoding 
algorithms \cite{kou,mac,mac1}. The construction of LDPC 
codes given in \cite{kou} was later generalized in different directions \cite{kam,tan,tan1,xu}, which resulted in 
several large classes of finite geometry LDPC codes. Codes based on the 
more general partial geometries were presented in \cite{hut,joh,li,van}.  In a recent 
paper \cite{diao}, it was shown that diverse classes of LDPC codes that appear 
in the literature are actually partial geometry codes although their construction 
methods were not based on geometric notions.

In this paper, aspects of partial geometries and their use in coding theory are considered. A characterization of a special category of partial geometries realized from an array of cyclic permutation matrices is given. The graph representation of a partial geometry is also described. Two new classes of partial geometries are constructed, one from prime fields and the other one from cyclic subgroups of prime orders of finite fields. 
New classes of {\it quasi-cyclic} (QC) LDPC codes are constructed based on these
new constructions of partial geometries. The problem of determining the 
sizes of trapping sets of such geometric codes, initiated in \cite{diao}, 
is continued here for both the partial geometries and the subclass of generalized quadrangles (GQs). 
Finally, since the eigenvalues of the adjacency matrices of all the 
geometric objects under investigation are known, the expansion properties 
of them can be determined and are summarized to consider the 
possibility that they may play a role in suggesting promising candidates for
LDPC codes for further investigation. Such candidates would have to be verified
by simulation.

\section{Definitions, Concepts and Structural Properties of Partial Geometries}\label{sect2}
 
Consider a system composed of a set ${\bf N}$ of $n$ points and a set ${\bf M}$ 
of $m$ lines where each line is a set of points. If a line $L$ contains a 
point $v$, we say that $v$ is on $L$ and that $L$ passes 
through $v$. If two points are on a 
line, then we say that the two points are adjacent and if two lines pass 
through the same point, then we say that the two lines intersect, otherwise 
they are parallel. The system composed of the sets ${\bf N}$ and ${\bf M}$ 
is a {\it partial geometry} \cite{bat,bos,cam} if 
the following conditions are satisfied for 
some fixed integers $\gamma \geq 2, \rho \geq 2$ and $\delta \geq 1$:
\begin{enumerate}
\item  Any two points are on at most one line; 
\item  Each point is on $\gamma$  lines; 
\item  Each line passes through $\rho$ points; 
\item  If a point $v$ is not on a line $L$, then there are exactly $\delta$
lines, each passing through $v$ and a point on $L$. 
\end{enumerate}
Such a partial geometry will be denoted by PaG$(\gamma , \rho , \delta )$
and $\gamma, \rho$ and $\delta$ are called the parameters of the partial 
geometry. The parameter $\delta$ is called the {\it connection number}
of the geometry.
 
A simple counting argument \cite{bat} shows that the partial geometry 
PaG$(\gamma , \rho , \delta )$ has exactly
\begin{equation}
\label{eq:par1}
n = \rho ((\rho -1)(\gamma -1) + \delta)/\delta
\end{equation}
points and
\begin{equation}
\label{eq:par2}
 m = \gamma((\rho -1)(\gamma -1) + \delta)/ \delta 
\end{equation}
lines.
 
If $v$ and $v^{\prime}$ are adjacent points, then there are exactly 
$\gamma \delta + \rho - \gamma - \delta -1$  points, such that each of these points 
is adjacent to both $v$ and $v^{\prime}$. On the other hand, if $v$ 
and $v^{\prime}$ are not adjacent, then there are 
exactly $\gamma \delta$ points, such that 
each of these points is adjacent to both $v$ and $v^{\prime}$. Each 
point $v$ is adjacent to $(\delta +1)(\rho -1)$ other points. Three 
adjacent non-colinear points form a triangle. It follows from the above 
adjacency property that two adjacent points $v$ and $v^{\prime}$
are on $\gamma \delta + \rho -\gamma - \delta -1$ triangles.
As will be discussed later, such a triangle in the (point-point) adjacency 
matrix of the graph of the (point-point) partial geometry leads to a
cycle of length 6 in the corresponding Tanner graph \cite{tann}, assuming $\delta >1$.
 
Well known examples of partial geometries are Euclidean and projective 
geometries \cite{car,man} over finite fields. If $\delta = \gamma -1$, the partial 
geometry PaG$(\gamma , \rho , \gamma -1 )$ is called a {\it net} 
which consists of $n = \rho^2$ points and $m = \gamma \rho$
lines. Each point $v$ not on a line $L$ is on a unique line which is parallel 
to $L$. Equivalently, there is a unique line $L^{\prime}$ that intersects $v$
and $L$. The set of $m = \gamma \rho$
lines in the net PaG$(\gamma , \rho , \gamma -1 )$ can be partitioned into $\gamma$
classes, each consisting of $\rho$ lines, such that all the lines in each class 
are parallel, any two lines in two different classes intersect, and 
each of the $n = \rho^2$ points is on a unique line in each class. 
These classes of lines are called {\it parallel bundles}. A two-dimensional 
Euclidean geometry (or affine geometry) \cite{man} is a net.

For every point $v$ in PaG$(\gamma , \rho , \gamma -1)$, there are 
exactly $\gamma$ lines that intersect at $v$, i.e.,  lines that pass 
through $v$. These lines are said to form an {\it intersecting bundle} at $v$, 
denoted by $\Delta (v)$. Notice that $v$ is on every line in $\Delta (v)$, 
there are exactly $\gamma ( \rho -1)$ points, each is on a unique 
line in $\Delta (v)$, and all the other $n - \gamma ( \rho -1) -1$ 
points in PaG$(\gamma , \rho , \gamma -1 )$ are not on any line in $\Delta (v)$.
If $\delta = \rho$, then every point in PaG$(\gamma , \rho , \rho )$ 
is adjacent to $v$ since every point is on a line in $\Delta (v)$. 
In this case, any two points in PaG$(\gamma , \rho , \rho )$
are connected by a line. Examples for which $\delta = \rho$ include
two-dimensional Euclidean and projective geometries (also called 
affine and projective planes) \cite{man}.
 
Denote the points and lines in PaG$(\gamma , \rho , \delta )$ 
by $v_0, v_1, \dots , v_{n-1}$
 and $L_0, L_1, \dots , L_{m-1}$, respectively. Then, 
${\bf N} = \{ v_0, v_1, \dots , v_{n-1} \}$ and ${\bf M} = \{ L_0, L_1, \dots , 
L_{m-1} \}$. Algebraically, a partial geometry 
PaG$(\gamma , \rho , \delta )$ with $n$ points and $m$ lines is commonly 
represented by a $m \times n$ matrix $\bH_{\text{PaG}} = [h_{i,j}]_{0 \leq i < m, ~
0 \leq j < n}$  with $0$ and $1$ 
entries whose rows and columns correspond to the lines and points of 
PaG$(\gamma , \rho , \delta )$, respectively. The rows are labeled from $0$
 to $m -1$ (or by the lines $\{L_0, L_1, \dots , L_{m-1} \}$ in this order) 
and the columns are 
labeled from $0$ to $n-1$ (or the points $\{ v_0, v_1, \dots , v_{n-1} \}$ in 
this order).  The entry $h_{i,j}$ is 1 (i.e., $h_{i,j} = 1$) if and only if 
the point $v_j$ labeled by $j$ (called the $j$-th point) is on the line 
$L_i$ labeled by $i$ (called the $i$-th line); otherwise, $h_{i,j} = 0$.  This 
matrix $\bH_{\text{PaG}}$ is called the {\it line-point adjacency matrix} of the partial 
geometry PaG$(\gamma , \rho , \delta )$ which shows the incidence relationship of 
the lines and points of PaG$(\gamma , \rho , \delta )$. The $i$-th row of 
$\bH_{\text{PaG}}$ is called the incidence vector of the $i$-th line $L_i$. The transpose 
$\bH_{\text{PaG}}^T$ of $\bH_{\text{PaG}}$ is also an adjacency matrix of PaG$(\gamma , \rho , \delta )$
called the {\it point-line adjacency matrix} of PaG$(\gamma , \rho , \delta )$.

Graphically, a partial geometry PaG$(\gamma , \rho , \delta )$ 
can be displayed by an adjacency graph $\mathcal{G} (\gamma , \rho , \delta )$ with its $n$ points 
represented by $n$ nodes or vertices, labeled by $v_0, v_1, \dots, v_{n-1}$. 
Two vertices $v_j$ and $v_k$ are connected by 
an edge $(v_j, v_k)$ if and only if they are on the 
same line. A line $L = \{ v_{j_0}, v_{j_1}, \dots , v_{j_{\rho-1}} \}$  in
PaG$(\gamma , \rho , \delta )$
is represented by a sequence of $\rho -1$ connected edges, 
$(v_{j_0}, v_{j_1}), (v_{j_1}, v_{j_2}), \dots , (v_{j_{\rho -2}}, v_{j_{\rho-1}})$. 
Each node $v_{j_k}$ in $\mathcal{G} (\gamma , \rho , \delta )$ has $\gamma$ edges 
incident with it and $\gamma$ is called the degree of $v_{j_k}$. It
is the number of lines in PaG$(\gamma , \rho , \delta )$
that intersect at the point $v_{j_k}$.

The partial geometry PaG$(\gamma , \rho , \delta )$ can also be represented by a 
{\it bipartite graph} \cite{str}, denoted by $\cG ({\bf V}, {\bf C})$, 
which is more commonly used in coding theory. It consists of two sets of nodes, 
denoted by ${\bf V}$ ( the set of {\it variable nodes} (VNs)) and ${\bf C}$ (the set of {\it check nodes} (CNs)), 
of size $n$ and $m$, respectively.  The $n$ nodes in ${\bf V}$, 
denoted by $v_0, v_1, \ldots , v_{n-1}$, represent the $n$ points  
in PaG$(\gamma , \rho , \delta )$ with $v_j$ representing the point $v_j$ and the $m$ nodes in ${\bf C}$, denoted 
by $c_0, c_1, \ldots , c_{m-1}$,  represent the $m$ 
lines in PaG$(\gamma , \rho , \delta )$  with $c_i$ representing the 
line $L_i$. A node $v_j$  in ${\bf V}$ is connected to a node $c_i$ in ${\bf C}$ 
if and only if the point $v_j$ in PaG$(\gamma , \rho , \delta )$ represented 
by the node $v_j$ is on the line $L_i$  in PaG$(\gamma , \rho , \delta )$ 
represented by the node $c_i$.  This bipartite graph 
$\cG ({\bf V},{\bf C})$ is called the {\it Tanner graph} \cite{tann}
associated with the line-point adjacency matrix $\bH_{\text{PaG}}$ of the partial
geometry PaG$(\gamma , \rho , \delta )$, commonly 
used in the study of LDPC codes in 
coding theory \cite{tann}. We call this bipartite graph a $( \gamma, \rho  )$- 
{\it biregular} bipartite graph since the VNs on the left are regular 
(have the same degree $\gamma$) and the CNs on the right are regular (of the same degree $\rho$).
The associated (or the adjacency)  matrix $\bH_{\text{PaG}}$ is used as the parity-check matrix of the code.

The girth of the $(\gamma , \rho )$-biregular bipartite graph 
$\cG( \bf V, \bf C )$ (defined as length 
of the shortest cycle) of the partial geometry PaG$(\gamma , \rho , \delta )$ for $\delta >1$, is 6 and there are 
$m \gamma (\gamma -1)(\delta -1) (\rho -1)/6$ cycles of length 6 
in $\cG( \bf V , \bf C )$ \cite{joh}. For $\delta =1$ the partial geometry is a
{\it generalized quadrangle} (GQ) \cite{pay}, 
and the corresponding graph will have girth of 8.
A simple modification of the argument of \cite{joh} to find the number of
cycles of length 8 is as follows. Consider a line $L$ and points $v_1, v_2$ on $L$.
Consider the intersecting bundles of lines, $\Delta (v_1 )$ and $\Delta (v_2 )$. 
Each set contains $(\gamma -1)$ lines other than $L$. Consider a point $v$ on 
a line $L_2$ through $v_2$ other than $L$ (i.e., in $\Delta (v_2 ) \backslash
\{L\}$) and a line $L_1$ through $v_1$ in $\Delta (v_1 ) \backslash \{L\}$. 
By the definition of a GQ, there is a unique line through $v$ that intersects
$L_1$. This is true for each point on the lines in $\Delta (v_2 )$, other than
$v_2$  and each line in $\Delta (v_1 )$ other than $L$. Hence, there are
$(\gamma -1)^2 (\rho -1 )$ distinct quadrangles in the GQ that contain $v_1, v_2$ 
as a side and $\binom{\rho}{2}$ ways of choosing the two points on a line. 
The argument is repeated for each of the
$m$ lines in the geometry and note that each quadrangle is repeated four times
in this count giving the number of quadrangles in the geometry (cycles of length 8
in the Tanner graph) as
\[
m (\rho -1) (\gamma -1)^2 \binom{\rho}{2}/4.
\]

\section{A Special Category of Partial Geometries} \label{sect3}
 
In this section, we present a special category of partial geometries which are 
specified by their line-point adjacency matrices. First, we prove 
the conditions for a matrix of a specific type to be the line-point
adjacency matrix of a partial geometry. Some structural 
properties of this category of partial geometries are then 
developed. In Sections \ref{sec:class} and \ref{sec:cyclegroup}, we will present two methods for constructing this category of partial geometries based on prime fields and cyclic subgroups
of prime orders of finite fields, respectively.

\subsection{Line-Point Adjacency Matrices}
Consider an $m \times n$ matrix $\bH$ with 0- and 1-entries for which two rows 
(or two columns) have at most one place where both have 1 entries. 
Such a constraint on rows and columns are called the {\it row and column constraint} (RC-constraint) \cite{kou}. The matrix $\bH$ is called an {\it RC-constrained matrix}. An RC-constrained matrix $\bH$ is {\it $( \gamma, \rho)$-biregular} if each of its $n$ columns has $\gamma$ 1-entries and each of its $m$ rows 
has $\rho$ 1-entries. The parameters $\gamma$ and $\rho$ are called the column 
and row weights of $\bH$, respectively. Suppose the rows of 
a $(\gamma , \rho )$-biregular matrix $\bH$ are labeled from 0 to $m-1$ and the columns 
from $0$ to $n-1$. The $\gamma$ rows of $\bH$ that have 1-entries at the $j$-th 
position with $0 \leq j < n$ are said to be attached to the $j$-th column. 
Likewise, the $\rho$ columns that have 1-entries in the $i$-th row are 
said to be attached to the $i$-th row. 
 
The following theorem gives the conditions that an RC-constrained $m \times n$
regular matrix $\bH$ with 0- and 1-entries is the line-point adjacency 
matrix of a partial geometry with $n$ points and $m$ lines.
 
\begin{thm}
\label{thm:one}
Let $\bH$ be an $m \times n$ RC-constrained $(\gamma , \rho )$-biregular 
matrix which is a $\gamma \times \rho$ array of $\gamma \times \gamma$ 
circulant permutation matrices (CPMs), where $m = \gamma^2$  
and $n = \gamma \rho$. Then, $\bH$ is the 
line-point adjacency matrix of a partial geometry PaG$(\gamma , \rho , \rho -1 )$ 
that has $n$ points corresponding to the columns of $\bH$ and $m$ lines 
corresponding to the rows of $\bH$.
\end{thm}

\begin{IEEEproof}
%\begin{proof}
Since $\bH$ is a $\gamma \times \rho$ array of $\gamma \times \gamma$ CPMs, 
it is a $\gamma^2 \times \gamma \rho$ $(\gamma , \rho)$-biregular matrix 
with constant column weight $\gamma$ and 
constant row weight $\rho$. It is clear that each point (corresponding to a 
column) is on $\gamma$ lines (corresponding to $\gamma$ rows) and each line 
(corresponding to a row) passes through $\rho$ points. The RC-constraint 
implies that any two points are on at most one line. 

It remains to show that if a point $v$ is not on a line $L$, then there are 
exactly $\rho -1$ lines that pass through $v$ and $\rho -1$ points on $L$. 
We group the $n$ points into $\rho$ sets of size $\gamma$ where the points 
in each set correspond to $\gamma$ consecutive columns in $\bH$, namely, the columns 
comprising a column-block of CPMs in $\bH$. Since every row has $\rho$ ones, then 
by adding all the rows attached to the column corresponding to the 
point $v$, where the sum is over the integers rather than over GF$(2)$, 
we obtain a vector, $\bz$, of length $n$ whose components as integers add 
up to $\rho \gamma$. Notice that the 
entry in the column corresponding to the point $v$ in $\bz$ is $\gamma$ while 
all other entries in columns corresponding to the $\gamma -1$ other 
points in the same set as $v$ are zeros since every row attached 
to the column corresponding to $v$ has zeros in all columns 
corresponding to other points in the same set as $v$ as every CPM 
has a single 1 in each row. Hence, all the $(\rho -1) \gamma$ components 
in $\bz$ other than those corresponding to $v$ and the $\gamma -1$ points 
in the same set as $v$ add up to $(\rho -1) \gamma$. Because of the
RC-constraint, all these components are at most equal to 1 implying 
that all of them are equal to 1. Hence, every column other than 
those corresponding to $v$ or other points in its set is attached 
to a unique row corresponding to a line passing through $v$ while 
every column corresponding to a point other than $v$ but in the 
same set as $v$ is not attached to any row corresponding to a line 
passing through $v$. Since $L$ passes through a single point in 
each set, it passes through $\rho -1$ points in sets other 
than that of $v$. Each of these points is on a line passing 
through $v$. The point on $L$ that is in the same set as $v$ is 
not adjacent to $v$. Hence, there are exactly $\rho -1$ lines 
that pass through $v$ and $\rho -1$ points on $L$. This proves that $\bH$ is 
the adjacency matrix of a partial geometry PaG$(\gamma , \rho , \rho -1 )$.
\end{IEEEproof}
%\end{proof}
 
The proof of the above theorem gives insight into 
the structure of the partial geometry PaG$(\gamma , \rho , \rho -1 )$
which is discussed further in following sections.

It is of interest to develop characterizations of partial geometries 
in the sense that if a combinatorial construction is developed,
what technique could be used to verify that it is a partial geometry?
The next discussion considers one such approach for a particular case 
where the adjacency matrix of the configuration is an array of CPMs.

Let $\bH$ be the $m \times n$ line-point adjacency matrix of a partial
geometry PaG$( \gamma, \rho, \delta)$. Suppose $m= m^{\prime}  \ell$ 
and $ n = n^{\prime} \ell$
and assume that $\bH$ is an  $m^{\prime} \times n^{\prime}$ array of CPMs
of order $\ell$, i.e., if $\bP$ is a generator of the cyclic  matrix group with the first (or the top) row the $\ell$-tuple $(0,1,0, \cdots , 0)$. The group of such $\ell \times \ell$
matrices is $\{ \bI , \bP , \dots , \bP^{\ell -1} \},  \bP^{\ell} = \bI$, where $\bI$ is an $\ell \times \ell$ identity matrix.

The matrix ${\bH}{\bH}^T$ is an $m \times m$ matrix of 0's and 1's whose $(i,j)$-th element
is 1 if the $i$-th line intersects the $j$-th line and 0 if the lines do not intersect 
and with diagonal elements $\rho$.
If the $i$-th column-block of CPMs of $\bH$ is multiplied by the $\ell \times \ell$
matrix $\bI + \bP^j$, $0 \leq i < n^{\prime}$ and $0 \leq j < \ell$, the effect is to
add to the circulant in a given row of circulants, the power of a 
nonunity permutation (circulant), i.e.,  it adds a 1 to each row of the circulant in the $i$-th
column-block  of circulants, not in the position of the existing 1. For $0 \leq i < n^{\prime}$ and $0 \leq j < \ell$, form the following $n^{\prime} \times n^{\prime}$ diagonal array of circulants of size $\ell \times \ell$:
\[
{\bD}^{(i,j)} = \text{diag} (\bI, \bI, \dots, \bI + \bP^j , \dots , \bI)
\]
where the matrix $\bI + \bP^j$ is in the $i$-th position of the array $\bD^{(i,j)}$. Then, the matrix
\begin{equation}
\label{eq:pag}
{\bG}^{(i,j)} = {\bH} {\bD}^{(i,j)} {\bH}^T ,~~ 0 \leq i < n^{\prime},~~ 0 \leq j < \ell 
\end{equation}
is an $m \times m$ matrix with $\delta$ 2's per row, corresponding to the
$\delta$ lines that intersect the line of that row with the extra 
point not on the line, diagonal elements of $\rho$
and all other elements 0 or 1. Note that the matrix ${\bH} {\bD}^{(i,j)}$
contains an extra point in each row, each row corresponding to a line
and a point not on the line. Conversely, if $\bH$ is an $m \times n$
binary matrix which is an  $m' \times n'$ array of  CPMs of order $\ell$ satisfying these conditions, it represents a partial geometry PaG$(\gamma , \rho , \delta )$. 
This is clear by construction. Thus, we have the following theorem that characterizes a binary array of  CPMs to be a line-point adjacency matrix of a partial geometry.

\begin{thm}
\label{thm:two}
Let $\bH$ be an $m \times n$ binary matrix written as an $m^{\prime} \times n^{\prime}$ 
array of CPMs of order $\ell$ whose associated matrices ${\bG}^{(i,j)}$
of (\ref{eq:pag}) satisfy the stated conditions. Then,  $\bH$ represents
a partial geometry PaG$( \gamma , \rho , \delta)$.
\end{thm}

The condition of the theorem does not allow zero matrix (ZM) of order $\ell$. The argument can be modified
to accommodate this case although the result is more complicated and not presented here.

In Sections \ref{sec:class} and \ref{sec:cyclegroup}, we will use Theorems \ref{thm:one} and \ref{thm:two} to construct two new classes of partial geometries whose line-point adjacency matrices are arrays of CPMs.

\subsection{Structural Properties}
\label{sec:struc}
 
Consider a partial geometry PaG$(\gamma , \rho , \rho -1 )$ as in the above Theorem \ref{thm:one}.
Notice that the line-point adjacency matrix $\bH$ consists of $\gamma$ row-blocks of CPMs and $\rho$ 
column-blocks of CPMs of size $\gamma \times \gamma$. We first group 
the $\gamma \rho$  points into $\rho$ sets, 
denoted by $\bN_0 , \bN_1 , \dots  , \bN_{\rho -1}$, each consisting 
of $\gamma$ points which correspond to $\gamma$ columns in a 
column-block of CPMs of the line-point adjacency matrix $\bH$.  
Since each row of a CPM has only one 1-entry, the $\rho$ points of a 
line $L$ are distributed in $\rho$ different sets, one point in each 
set.  For $0 \leq k < \rho$, a point $v_k$ in the set $\bN_j$ that 
is not on a line $L$ in PaG$(\gamma , \rho , \rho -1 )$ is not adjacent 
to the point on $L$ in the 
same set $\bN_j$ but is adjacent to each of the other $\rho -1$
points on $L$ that are in the $\rho -1$  sets other than $\bN_j$.
 
The $\gamma^2$ lines corresponding to the $\gamma^2$ rows of the line-point 
adjacency matrix $\bH$ can be grouped into $\gamma$ bundles, denoted 
by $\bM_0, \bM_1, \dots , \bM_{\gamma -1}$, each consisting of $\gamma$ 
lines which correspond 
to the $\gamma$ rows in a row-block of CPMs in $\bH$.  Since any two rows 
in a row-block of CPMs in $\bH$ do not have any position where they 
both have 1-entries, the $\gamma$ lines in each bundle $\bM_i$ are parallel 
lines. Hence, each bundle $\bM_i$ corresponds to a parallel bundle 
of lines in PaG$(\gamma , \rho , \rho -1 )$.

Let $\bH_0, \bH_1,\dots , \bH_{\gamma -1} $ denote the $\gamma$ 
row-blocks of $\bH$. For $0 \leq  i < \gamma$
and $0 \leq k < \gamma$, let $\bh_{i,k} = (\bh_{i,k,0},  \bh_{i,k,1} , 
\dots , \bh_{i,k, \rho-1} ) $
be the $k$-th row in the $i$-th row-block $\bH_i$ which consists of $\rho$
sections, $\bh_{i,k,0},  \bh_{i,k,1} , \dots ,  \bh_{i,k,\rho -1}$, 
each consisting of $\gamma$ components. Each section is the $k$-th 
row of a CPM in $\bH_i$. If we cyclically 
shift all the $\rho$ sections of $\bh_{i,k}$ simultaneously {\it one 
place to the right within the sections}, we obtain the $(k + 1)$-th row 
$\bh_{i,k+1}$ of $\bH_i$ which also consists of $\rho$ sections    
and each section is the 
$(k + 1)$-th row of a CPM in $\bH_i$. The above cyclic-shift within each 
section of $\bh_{i,k}$ is referred to as {\it section-wise cyclic-shift} of 
the row $\bh_{i,k}$.  For $k = \gamma - 1$, the section-wise cyclic-shift of 
$\bh_{i, \gamma -1}$  results in the top row $\bh_{i,0}$ of $\bH_i$.  
Consequently, all the rows of the $i$-th row-block $\bH_i$ can 
be obtained by section-wise cyclically shifting the top row $\bh_{i,0}$
 of $\bH_i$  $\gamma -1$ times. 

Let ${\bH}^{\ast}_0$ be an $\gamma \times \gamma \rho $ matrix over GF$(2)$ 
which consists of the top rows $\bh_{0,0},  \bh_{1,0}, \dots ,  \bh_{\gamma -1,0}$
of the $\gamma$ row-blocks 
$\bH_0, \bH_1, \dots , \bH_{\gamma -1}$ of the line-point adjacency 
matrix $\bH$ of PaG$(\gamma , \rho , \rho -1 )$. Then, it follows 
from the section-wise 
cyclic structure of $\bH$, the entire array $\bH$ can be obtained by 
section-wise cyclically shifting ${\bH}^{\star}_0$  $\gamma -1$ times. 
This structural property is referred to as {\it QC} (or {\it section-wise cyclic}) {\it structure}.  The QC-structure 
of $\bH$ implies that the incidence vectors of the $\gamma^2$ lines in 
the partial geometry PaG$(\gamma , \rho , \rho -1 )$ which correspond to 
the $\gamma^2$ rows of $\bH$ can be generated by the incidence vectors of 
the $\gamma$ lines which correspond to the $\gamma$ rows in ${\bH}^{\ast}_0$.  
That is to say that the lines corresponding to the rows of ${\bH}^{\ast}_0$
and the points on these lines specify the partial 
geometry PaG$(\gamma , \rho , \rho -1 )$.
 
Due to the QC-structure of $\bH$, we call the partial geometry 
PaG$(\gamma , \rho , \rho -1 )$ a {\it quasi-cyclic partial geometry}, denoted by QC-PaG$(\gamma , \rho , \rho -1 )$.
 
In the following, we give an example to demonstrate that a QC-PaG does exist.
\vspace{.1in}
 
\begin{exmp}\label{eg1}
Consider the $3 \times 3$ array $\bH$ of CPMs of size $3 \times 3$ given in the equation below: 

\begin{equation}
\label{eq:mat1}
\bH = \;
\begin{bmatrix}
1 & 0 & 0 & 1 & 0 & 0 & 1 & 0 & 0\\
0 & 1 & 0 & 0 & 1 & 0 & 0 & 1 & 0\\
0 & 0 & 1 & 0 & 0 & 1 & 0 & 0 & 1\\
1 & 0 & 0 & 0 & 1 & 0 & 0 & 0 & 1\\
0 & 1 & 0 & 0 & 0 & 1 & 1 & 0 & 0 \\
0 & 0 & 1 & 1 & 0 & 0 & 0 & 1 & 0\\
1 & 0 & 0 & 0 & 0 & 1 & 0 & 1 & 0\\
0 & 1 & 0 & 1 & 0 & 0 & 0 & 0 & 1\\
0 & 0 & 1 & 0 & 1 & 0 & 1 & 0 &0
\end{bmatrix}.
\end{equation}

It is a $9 \times 9$ matrix over GF$(2)$.  By checking, 
we find that it satisfies the RC-constraint. Hence, it follows from Theorem \ref{thm:one} that $\bH$  is a line-point 
adjacency matrix of a PaG$(3, 3, 2)$ with 9 points, 9 
lines and connection number 2. Each line consists of 3 points and 
each point is on 3 lines.  The lines in PaG$(3, 3, 2)$ can be 
partitioned into 3 parallel bundles, each consisting of 3 parallel 
lines. Suppose we use $v_0, v_1, v_2, v_3, v_4, v_5, v_6, v_7, v_8$  to denote the 
9 points in PaG$(3, 3, 2)$ which correspond to the 9 columns of $\bH$. 
Then, the 9 lines in PaG$(3, 3, 2)$ are: 
\vspace{.1in}

\[
\begin{array}{rl}
L_0 & = \{v_0, v_3, v_6 \}, \\
L_1 & = \{v_1, v_4, v_7 \},  \\
L_2 & = \{v_2, v_5, v_8 \}, \\
L_3 & = \{v_0, v_4, v_8 \}, \\
L_4 & = \{v_1, v_5, v_6 \}, \\
L_5 & = \{v_2, v_3, v_7 \}, \\
L_6 & = \{v_0, v_5, v_7 \}, \\
L_7 & = \{v_1, v_3, v_8 \}, \\
L_8 & = \{v_2, v_4, v_6 \}.  
\end{array}
\]
\vspace{.1in}

Group the points in PaG$(3, 3, 2)$ into 3 disjoint sets, $\bN_0, \bN_1$ and $\bN_2$, 
each consisting of 3 points. For $0 \leq j < 3$, the 3 points in $\bN_j$ 
correspond to the 3 columns in the $j$-th column-block of $\bH$. 
Then, $\bN_0 = \{ v_0, v_1, v_2\}, \bN_1 = \{v_3, v_4, v_5 \} ~\text{and}~ \bN_2 = \{ v_6, v_7, v_8 \}$.  
Consider the line $L_3 = \{ v_0, v_4, v_8 \}$.  The point $v_5$ is not on $L_3$. 
By checking, we find that this point is adjacent to the points $v_0$
and $v_8$. Notice that the point $v_5$ is in the set $\bN_1$ but its two adjacent 
points $v_0$ and $v_8$ are in two separate sets $\bN_0$ and $\bN_2$. The graphical 
representation of PaG$(3, 3, 2)$ is shown in Fig. \ref{fig:repsn} and the Tanner 
graph associated to the line-point adjacency matrix $\bH$ is shown 
in Fig. \ref{fig:tann}. The girth of the graph is 6. 
%A cycle of length 6 is shown in bold lines.

\begin{center}
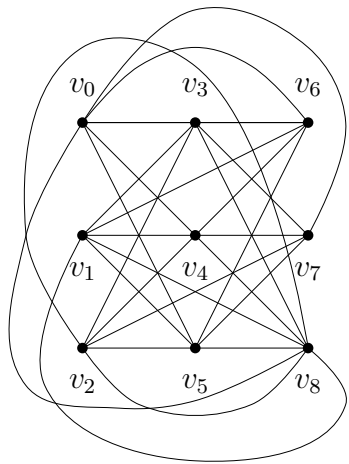
\begin{figure}[!h]
\begin{center}
\begin{pspicture}(-2,-1)(8,5.5)
%\psline(0,0)(10,5)
\psset{unit=.5cm}
%\psset{linewidth=.01cm}
\pslinewidth=.3pt
\qdisk(2,2){2pt}
\qdisk(2,5){2pt}
\qdisk(2,8){2pt}
\qdisk(5,2){2pt}
\qdisk(5,5){2pt}
\qdisk(5,8){2pt}
\qdisk(8,2){2pt}
\qdisk(8,5){2pt}
\qdisk(8,8){2pt}
\uput{10pt}[90](2,8){$v_0$}
\uput{10pt}[90](5,8){$v_3$}
\uput{10pt}[90](8,8){$v_6$}
\uput{10pt}[270](2,5){$v_1$}
\uput{10pt}[270](5,5){$v_4$}
\uput{10pt}[270](8,5){$v_7$}
\uput{10pt}[270](2,2){$v_2$}
\uput{10pt}[270](5,2){$v_5$}
\uput{10pt}[270](8,2){$v_8$}
\psline(2,8)(8,8)
\psline(2,8)(8,2)
\psline(2,8)(5,2)
\psline(5,8)(2,5)
\psline(5,8)(2,2)
\psline(5,8)(8,2)
\psline(5,8)(8,5)
\psline(8,8)(2,5)
\psline(8,8)(2,8)
\psline(8,8)(2,2)
\psline(8,8)(5,2)
%%%%%%%%%%%%%%%%%%%%%%%%
\psline(2,5)(5,2)
\psline(2,5)(8,2)
\psline(2,5)(8,5)
%%%%%%%%%%%%%%%%%%%%%%%
\psline(8,5)(2,2)
\psline(8,5)(5,2)
\psline(2,2)(8,2)
%%%%%%%%%%%%%%%%%%%%%
\pscurve(2,8)(5,11)(8,10)(9,8)(8,5)
\pscurve(2,8)(3.5,9.5)(5,10)(6.5,9.5)(8,8)
\pscurve(2,8)(.5,5)(.5,1)(3,.4)(5,0.6)(8,2)
\pscurve(8,2)(5,9.5)(2,10)(0.5,5)(2,2)
\pscurve(2,5)(1,1)(5,-1)(9,.5)(8,2)
\pscurve(2,2)(3.5,0.5)(6.5,0.5)(8,2)
\end{pspicture}
\caption{A graphical representation of the partial geometry PaG$(3, 3, 2)$ specified by its line-point adjacency matrix given by (\ref{eq:mat1}) in Example \ref{eg1}.}
\label{fig:repsn}
\end{center}
\end{figure}
\end{center}
\vspace{-.2in}

\begin{center}
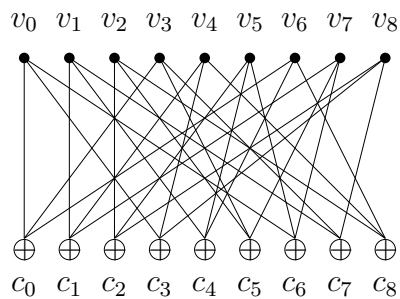
\begin{figure}[!h]
\begin{center}
\begin{pspicture}(-1,0)(9,5)
\psset{unit=.6cm}
%\psset{linewidth=.01cm}
\pslinewidth=.3pt
\qdisk(1,6){2pt}
\qdisk(2,6){2pt}
\qdisk(3,6){2pt}
\qdisk(4,6){2pt}
\qdisk(5,6){2pt}
\qdisk(6,6){2pt}
\qdisk(7,6){2pt}
\qdisk(8,6){2pt}
\qdisk(9,6){2pt}
\pscircle(1.0,1.75){4pt}
\pscircle(2.0,1.75){4pt}
\pscircle(3,1.750){4pt}
\pscircle(4,1.750){4pt}
\pscircle(5,1.750){4pt}
\pscircle(6,1.750){4pt}
\pscircle(7,1.750){4pt}
\pscircle(8,1.750){4pt}
\pscircle(9,1.750){4pt}
\uput{10pt}[90](1,6){$v_0$}
\uput{10pt}[90](2,6){$v_1$}
\uput{10pt}[90](3,6){$v_2$}
\uput{10pt}[90](4,6){$v_3$}
\uput{10pt}[90](5,6){$v_4$}
\uput{10pt}[90](6,6){$v_5$}
\uput{10pt}[90](7,6){$v_6$}
\uput{10pt}[90](8,6){$v_7$}
\uput{10pt}[90](9,6){$v_8$}
\uput{15pt}[270](1,2){$c_0$}
\uput{15pt}[270](2,2){$c_1$}
\uput{15pt}[270](3,2){$c_2$}
\uput{15pt}[270](4,2){$c_3$}
\uput{15pt}[270](5,2){$c_4$}
\uput{15pt}[270](6,2){$c_5$}
\uput{15pt}[270](7,2){$c_6$}
\uput{15pt}[270](8,2){$c_7$}
\uput{15pt}[270](9,2){$c_8$}
\uput{0pt}[0](.75,1.75){$+$}
\uput{0pt}[0](1.75,1.75){$+$}
\uput{0pt}[0](2.75,1.75){$+$}
\uput{0pt}[0](3.75,1.75){$+$}
\uput{0pt}[0](4.75,1.75){$+$}
\uput{0pt}[0](5.75,1.75){$+$}
\uput{0pt}[0](6.75,1.75){$+$}
\uput{0pt}[0](7.75,1.75){$+$}
\uput{0pt}[0](8.75,1.75){$+$}
\psline(1,6)(1,2)
\psline(1,6)(4,2)
\psline(1,6)(7,2)
\psline(2,6)(2,2)
\psline(2,6)(5,2)
\psline(2,6)(8,2)
%%%%%%%%%%%%%%%%%%%%%%%%%%
\psline(3,6)(3,2)
\psline(3,6)(6,2)
\psline(3,6)(9,2)
\psline(4,6)(1,2)
\psline(4,6)(6,2)
\psline(4,6)(8,2)
%%%%%%%%%%%%%%%%%%%%%%%
\psline(5,6)(2,2)
\psline(5,6)(4,2)
\psline(5,6)(9,2)
\psline(6,6)(3,2)
\psline(6,6)(5,2)
\psline(6,6)(7,2)
%%%%%%%%%%%%%%%%%%%%%
\psline(7,6)(1,2)
\psline(7,6)(5,2)
\psline(7,6)(9,2)
\psline(8,6)(2,2)
\psline(8,6)(6,2)
\psline(8,6)(7,2)
\psline(9,6)(3,2)
\psline(9,6)(4,2)
\psline(9,6)(8,2)
\end{pspicture}
\end{center}
\caption{The Tanner graph associated with the line-point adjacency matrix of PaG$(3, 3, 2)$ given by (\ref{eq:mat1}) in Example \ref{eg1}.}
\label{fig:tann}
\end{figure}
\end{center}
\vspace{.1in}
\end{exmp}
\vspace{-.1in}

\section{Protograph Representation of a QC-PaG} \label{sec:prot}
 
As shown in Section \ref{sect2}, a partial geometry with $n$ points and $m$ lines
can be represented by a bipartite graph with $n$ VNs which represent the $n$ points of the geometry and
$m$ CNs which represent the $m$ lines of the geometry. A VN $v_j$ is connected to a CN $c_i$ by an edge
$(i, j)$ if and only if the point represented by the VN $v_j$ is on the line represented by the CN $c_i$.
For a partial geometry with a large set of points and a large set of lines, the bipartite graph representation
of the geometry would be very large and complex.  However, the QC-PaG$(\gamma , \rho , \rho -1 )$ characterized by Theorem 1 can be effectively represented by a much smaller bipartite graph based on its QC-structure.

Recall that the line-point adjacency matrix $\bH$ of the QC-PaG$(\gamma , \rho , \rho -1 )$ is a binary $\gamma \times \rho$ array of CPMs of order $\gamma$ of the following form:

\begin{equation}
\label{eqn:hmat}
\begin{array}{rl}
{\bH} & = \;
\begin{bmatrix}
{\bH_0}\\
{\bH_1}\\
\vdots \\
{\bH_{\gamma -1}}
\end{bmatrix} \\
& = \;
\begin{bmatrix}
\bA_{0,0} & \bA_{0,1} & \cdots & \bA_{0,\rho -1} \\
\bA_{1,0} & \bA_{1,1} & \cdots & \bA_{1,\rho -1} \\
\vdots & \vdots & \cdots & \vdots \\
\bA_{\gamma -1,0} &\bA_{\gamma -1,1} & \cdots & \bA_{\gamma -1,\rho -1}\\
\end{bmatrix}
\end{array}.
\end{equation}

The array $\bH = [\bA_{i,j}]_{0 \le i < \gamma, 0 \le j < \rho}$ consists of $\rho$ column-blocks of CPMs, denoted by $\bN_0, \bN_1, \ldots,\linebreak \bN_{\rho-1}$, and $\gamma$ row-blocks of CPMs,
denoted by $\bH_0, \bH_1,\ldots, \bH_{\gamma-1}$. For $0 \le i < \gamma$ and $0 \le j < \rho$, each column-block $\bN_j$ consists of $\gamma$ consecutive
columns of $\bH$ and each row-block $\bH_i$ consists of $\gamma$ consecutive rows of $\bH$. The $\gamma$ columns of the $j$-th column-block $\bN_j$
correspond to $\gamma$ points of the QC-PaG$(\gamma , \rho , \rho -1 )$ and the $\gamma$ rows of the $i$-th row-block $\bH_i$ correspond to $\gamma$ lines of the QC-PaG$(\gamma , \rho , \rho -1 )$.

For $0 \le j < \rho$, let $\Phi_j$ be the set of $\gamma$ points which correspond to the $\gamma$ columns of the $j$-th column-block $\bN_j$ of $\bH$. For $0 \le i < \gamma$, let $\Omega_i$ be the set of lines which correspond to the $\gamma$ rows of the $i$-th row-block $\bH_i$ of $\bH$. In forming the Tanner graph $\mathcal{G}_{\text{PaG}}$ of the QC-PaG$(\gamma , \rho , \rho -1 )$, the $\gamma$ points in $\Phi_j$ are represented by $\gamma$ VNs and the $\gamma$ lines in $\Omega_i$ are represented by $\gamma$ CNs. Hereafter, we use points and VNs interchangeably, lines and CNs interchangeably. The $\gamma$ VNs in $\Phi_j$ are called type-$j$ VNs and the $\gamma$ CNs in $\Omega_i$ are called type-$i$ CNs. From the QC-structure of $\bH$, we see that a type-$j$ VN can only be connected to a type-$i$ CN and vise versa. In forming the Tanner graph $\mathcal{G}_{\text{PaG}}$ of the QC-PaG$(\gamma , \rho , \rho -1 )$, the $\gamma$ type-$j$ VNs in $\Phi_j$ are connected to the $\gamma$ type-$i$ CNs in $\Omega_i$ based on the $\gamma$ 1-entires in the CPM $\bA_{i,j}$ and vise versa. If we label the columns and rows of a CPM $\bA_{i,j}$ in $\bH$ from 0 to $\gamma - 1$, then $\bA_{i,j}$ is uniquely specified by the location of the single 1-entry of its top row, called the {\it generator}.  If the single 1-entry of the top row of $\bA_{i,j}$ locates at the position $k_{i,j}$, $0 \le k_{i,j} < \gamma$, then we use ($k_{i,j}$) to specify the CPM $\bA_{i,j}$.

With all the terms defined above, we now construct a bipartite graph, denoted by $\mathcal{G}_{\text{proto,PaG}}$, with $\rho$ VNs and $\gamma$ CNs as shown in Fig. \ref{fig:proto}.  The $\rho$ VNs of $\mathcal{G}_{\text{proto, PaG}}$ represent the $\rho$ clusters $\Phi_0,  \Phi_1, \ldots, \Phi_{\rho-1}$ of VNs in $\mathcal{G}_{\text{PaG}}$ and the $\gamma$ CNs in $\mathcal{G}_{\text{proto,PaG}}$  represent the $\gamma$ clusters $\Omega_0, \Omega_1, \ldots, \Omega_{\gamma-1}$ of CNs in $\mathcal{G}_{\text{PaG}}$.  In $\mathcal{G}_{\text{proto,PaG}}$, the VN $\Phi_j$ is connected to the CN  $\Omega_i$  by an edge labeled by ($k_{i,j}$) which is the location of the single 1-entry of the generator (or top row) of the CPM $\bA_{i,j}$ in the line-point adjacency matrix $\bH$ of the QC-PaG$(\gamma , \rho , \rho -1 )$. The bipartite graph $\mathcal{G}_{\text{proto,PaG}}$   contains all the structural information of the QC-PaG$(\gamma , \rho , \rho -1 )$. The size of the bipartite graph $\mathcal{G}_{\text{proto,PaG}}$ is smaller than the size of the bipartite graph $\mathcal{G}_{\text{PaG}}$ by a factor $\gamma$.  This bipartite graph $\mathcal{G}_{\text{proto,PaG}}$ is called the {\it protograph} of the QC-PaG$(\gamma , \rho , \rho -1 )$. Basically, the protograph $\mathcal{G}_{\text{proto,PaG}}$ of the QC-PaG$(\gamma , \rho , \rho -1 )$ consists of $\rho$ super-VNs, $\Phi_0,  \Phi_1, \ldots, \Phi_{\rho-1}$, and $\gamma$ super-CNs, $\Omega_0, \Omega_1, \ldots, \Omega_{\gamma-1}$. The Tanner graph $\mathcal{G}_{\text{PaG}}$ of the QC-PaG$(\gamma , \rho , \rho -1 )$ is simply an expansion of the protograph $\mathcal{G}_{\text{proto,PaG}}$ of the QC-PaG$(\gamma , \rho , \rho -1 )$.

\begin{center}
\begin{figure}[!h]
\begin{center}
\begin{pspicture}(-1,-0.4)(11.5,2.5)
\psset{unit=.5cm}
%\psset{linewidth=.01cm}
\pslinewidth=.3pt
\qdisk(0,5){2pt}
\qdisk(3,5){2pt}
\qdisk(7,5){2pt}
\qdisk(12,5){2pt}

\pscircle(2,0){5pt}
\pscircle(6,0){5pt}
\pscircle(10,0){5pt}

\uput{10pt}[90](0,5){$\Phi_0$}

\uput{10pt}[90](3,5){$\Phi_1$}
\uput{10pt}[90](7,5){$\Phi_j$}
\uput{10pt}[90](12,5){$\Phi_{\rho-1}$}
\uput{10pt}[0](4,5){$\ldots$}
\uput{10pt}[0](8,5){$\ldots$}

\uput{10pt}[270](2,0){$\Omega_0$}
\uput{10pt}[270](6,0){$\Omega_i$}
\uput{10pt}[270](10,0){$\Omega_{\gamma-1}$}
\uput{10pt}[0](3,0){$\ldots$}
\uput{10pt}[0](7,0){$\ldots$}

\uput{0pt}[0](1.65,0){$+$}
\uput{0pt}[0](5.65,0){$+$}
\uput{0pt}[0](9.65,0){$+$}

%%%%%%%%%%%%%%%%%%%%%%%%%%%%
\psline(0,5)(2,0.35)
\psline(0,5)(1.5,4.5)
\psline(0,5)(6,0.35)

\psline(3,5)(2,4.5)
\psline(3,5)(3,4.1)
\psline(3,5)(3.6,4.4)

\psline(7,5)(2,0.35)
\psline(7,5)(6,0.35)
\psline(7,5)(10,0.35)
%\psline(7,5)(7,4.5)

\psline(12,5)(10,0.35)
\psline(12,5)(2,0.35)
\psline(12,5)(11,4)

%%%%%%%%%%%%%%%%%%%%%%%
\psline(2,0.35)(2.1,1.2)

\psline(6,0.35)(6.6,0.6)
\psline(10,0.35)(9.4,0.6)
%\psline(10,0.35)(10.5,0.5)

\uput{1pt}[0](-1,2.5){($k_{0,0}$)}
\uput{1pt}[0](2.5,3.3){($k_{i,0}$)}
\uput{1pt}[0](3.8,4.2){($k_{0,j}$)}
\uput{1pt}[0](6.5,2){($k_{i,j}$)}
\uput{1pt}[0](7.5,4.6){($k_{\gamma-1,j}$)}

\uput{1pt}[0](8.4,3){($k_{0,\rho-1}$)}
\uput{1pt}[0](11.5,3.5){($k_{\gamma-1,\rho-1}$)}

\end{pspicture}
\caption{The protograph of the QC-PaG$(\gamma , \rho , \rho -1 )$.}
\label{fig:proto}
\end{center}
\end{figure}
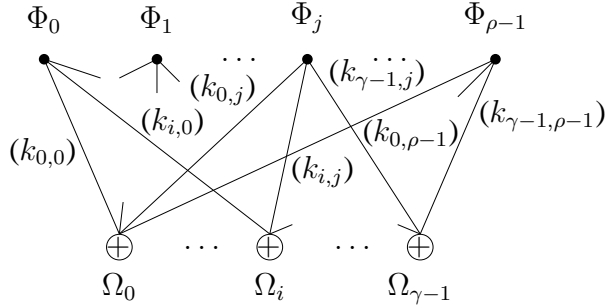
\end{center}
\vspace{.1in}

\begin{exmp} \label{eg2}
Consider the QC-PaG$(3,3,2)$ given in Example \ref{eg1}. We group the 9 points of the geometry into three disjoint sets, $\Phi_0 = \{v_0, v_1, v_2\}$, $\Phi_1 = \{v_3, v_4, v_5\}$, $\Phi_2= \{v_6, v_7, v_8\}$,
and the 9 lines of the geometry into three disjoint sets, $\Omega_0 = \{L_0, L_1,L_2\}$, $\Omega_1= \{L_3, L_4,L_5\}$, $\Omega_2= \{L_6, L_7,L_8\}$.
Using  $\Phi_0$,  $\Phi_1$,  $\Phi_2$ as the super-VNs and $\Omega_0$, $\Omega_1$, $\Omega_2$ as the super-CNs, we form the protograph  $\mathcal{G}_{proto,PaG}$ of the QC-PaG$(3,3,2)$ 
as shown in Fig. \ref{fig:proto2}. Each edge is labeled by the position of the single 1-entry of the generator of each CPM which specifies the connections between the constituent VNs of a super-VN $\Phi_j$ in  $\mathcal{G}_{proto,PaG}$ and the constituent CNs of a super-CN $\Omega_i$ in  $\mathcal{G}_{\text{proto,PaG}}$.

\begin{center}
\begin{figure}[!h]
\begin{center}
\begin{pspicture}(-1,-0.5)(11.5,2.5)
\psset{unit=.5cm}
%\psset{linewidth=.01cm}
\pslinewidth=.3pt
\qdisk(2,5){2pt}
\qdisk(6,5){2pt}
\qdisk(10,5){2pt}

\pscircle(2,0){5pt}
\pscircle(6,0){5pt}
\pscircle(10,0){5pt}

\uput{10pt}[90](2,5){$\Phi_0$}
\uput{10pt}[90](6,5){$\Phi_1$}
\uput{10pt}[90](10,5){$\Phi_2$}

\uput{10pt}[270](2,0){$\Omega_0$}
\uput{10pt}[270](6,0){$\Omega_1$}
\uput{10pt}[270](10,0){$\Omega_2$}

\uput{0pt}[0](1.65,0){$+$}
\uput{0pt}[0](5.65,0){$+$}
\uput{0pt}[0](9.65,0){$+$}

%%%%%%%%%%%%%%%%%%%%%%%%%%%%
\psline(2,5)(2,0.35)
\psline(2,5)(6,0.35)
\psline(2,5)(10,0.35)

\psline(6,5)(2,0.35)
\psline(6,5)(6,0.35)
\psline(6,5)(10,0.35)

\psline(10,5)(2,0.35)
\psline(10,5)(6,0.35)
\psline(10,5)(10,0.35)

%%%%%%%%%%%%%%%%%%%%%%%
\uput{0pt}[0](0.8,3){($0$)}
\uput{0pt}[0](2.2,3){($0$)}
\uput{0pt}[0](2.8,4.9){($0$)}

\uput{0pt}[0](4.3,4.5){($0$)}
\uput{0pt}[0](5,3.5){($1$)}
\uput{0pt}[0](6.8,4.5){($2$)}

\uput{0pt}[0](8,4.8){($0$)}
\uput{0pt}[0](9,3){($1$)}
\uput{0pt}[0](10.2,4){($2$)}

\end{pspicture}
\caption{The protograph of the QC-PaG$(3,3,2)$ given in Example \ref{eg2}.}
\label{fig:proto2}
\end{center}
\end{figure}
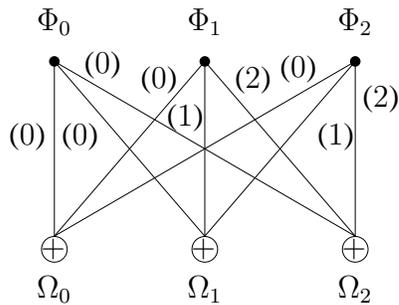
\end{center}
%\vspace{.1in}

\end{exmp}

LDPC codes can be constructed based on relatively small protographs \cite{abu, div, div1, tho}. These LDPC codes are called {\it protograph LDPC codes}. In constructing such a code, we take $\gamma$ copies of a chosen protograph.  Then, we permute the edges of the individual copies and connect the copies into a large Tanner graph.  The null space of the adjacency matrix of the Tanner graph gives a protograph LDPC code. Constructions of protograph LDPC codes in most of the literature are pseudo-random and require computer aid to perform the connections of the copies of the chosen protograph.

From the graph point of view, the QC-LDPC code constructed based on the QC-PaG$(\gamma , \rho , \rho -1 )$ may be regarded as a protograph QC-LDPC code. Based on the description of the QC-PaG$(\gamma , \rho , \rho -1 )$, we first construct its protograph $\mathcal{G}_{\text{proto,PaG}}$.  Then, we take $\gamma$ copies of $\mathcal{G}_{\text{proto,PaG}}$, permute the edges of the copies and connect the copies to form the Tanner graph $\mathcal{G}_{\text{PaG}}$ of the QC-PaG$(\gamma , \rho , \rho -1 )$. The null space of the adjacency matrix of the Tanner graph  $\mathcal{G}_{\text{PaG}}$ of the QC-PaG$(\gamma , \rho , \rho -1 )$ gives a protograph QC-LDPC code. In Sections \ref{sec:class} and \ref{sec:cyclegroup}, we will present two classes of QC-PaGs. The protograph of a QC-PaG from each of these two classes can be constructed directly from the description of its base matrix.

\section{A Class of QC-PaGs Constructed Based on Prime Fields} \label{sec:class}

In this section, we present a class of QC-PaGs based on prime fields whose line-point adjacency matrices are arrays of CPMs.

\subsection{Construction}
\label{sec:const}

Let $p$ be a prime and GF$(p)$ be a prime field which consists of the following 
$p$ elements: $0, 1, 2, . . . , p - 1$. We represent each element $i$ in GF$(p)$ 
by a $p \times p$ CPM, denoted by $\bQ(i)$, with columns and rows labeled from 
$0$ to $p - 1$, whose generator has 
its single 1-component at the location $i$. This representation is 
one-to-one. For $i = 0,~\bQ(0)$ is a $p \times p$ identity matrix. $\bQ (i)$ is referred 
to as the \emph{CPM-dispersion} of the element $i$ in GF$(p)$.

Form the following $p \times p$ matrix $\bB_{p}$ over GF$(p)$ with columns and 
rows labeled from $0$ to $p - 1$:
\vspace{.1in}

\begin{equation}
\label{eq:base}
\bB_{p} = \;
\begin{bmatrix}
0 \cdot 0 & 0 \cdot 1  & \cdots & 0 \cdot (p-1) \\
1 \cdot 0 & 1 \cdot 1  & \cdots & 1 \cdot (p-1) \\
2 \cdot 0 & 2 \cdot 1  & \cdots & 2 \cdot (p-1) \\
\vdots & \vdots  & \cdots & \vdots \\
(p-1) \cdot 0 & (p-1) \cdot 1 & \cdots & (p-1) \cdot (p-1) 
\end{bmatrix}
,
\end{equation}
\vspace{.1in}

\noindent
i.e., $\bB_{p} = [ b_{ij} ]_{0 \leq i,j < p-1}$ with $b_{ij} = i j$,
where the multiplication of two elements in GF$(p)$ is carried out under modulo $p$.
 
The matrix $\bB_{p}$ has the following structural properties: 
\begin{enumerate}
\item All the entries in the 0-th row and column  are zeros; 
\item All the entries in any row or column, other than the 0-th row and column,
are different and contain all the elements of GF$(p)$; 
\item Two different rows or columns have the 0 element of GF$(p)$ 
in common at the 0-th row or column and differ in all the other $p - 1$ 
positions;  
\item  The $i$-th column of $\bB_{p}$ is identical to the transpose 
of $i$-th row of $\bB_{p}$ for $0 \leq i <p$. 
\end{enumerate}
The last property implies that the transpose ${\bB}_{p}^T$ of $\bB_{p}$ 
is identical to $\bB_{p}$, i.e., ${\bB_{p}}^T = \bB_{p}$.
 
If each entry in $\bB_{p}$ is dispersed into its corresponding $p \times p$ CPM, we 
obtain a $p \times p$  array $\bH_{p}$ of CPMs of size $p \times p$.  
It is a $p^2 \times p^2$ matrix over GF$(2)$.
Each row has weight $p$ and each column has weight $p$. Based on the structural properties of $\bB_{p}$, we can readily 
see that $\bH_{p}$ satisfies the RC-constraint. Hence, it follows from 
Theorem \ref{thm:one} that $\bH_{p}$ is the line-point adjacency matrix of a QC-PaG$_{p}(p, p, p -1)$ with columns corresponding to the points and rows corresponding to the lines in the QC-PaG$_{p}(p, p, p - 1)$.  
Equally, Theorem \ref{thm:two} could have been applied. The partial geometry
QC-PaG$_{p}(p, p, p - 1)$ consists of $n = p^2$ points and $m = p^2$ lines. 
Each line in the QC-PaG$_{p}(p, p, p - 1)$ consists of $p$ points and each point 
is on $p$ lines. A point $v$ that is not on a line $L$ is connected 
to $p - 1$ points on $L$ by lines, i.e., the connection number of a point 
in the QC-PaG$_{p}(p, p, p - 1)$ is $p - 1$.  The array $\bH_{p}$ is called the 
{\it CPM-dispersion} of $\bB_{p}$ and $\bB_{p}$ is called the {\it base matrix} for the 
construction of the QC-PaG$_{p}(p, p, p - 1)$. The subscript ``$p$'' of the QC-PaG$_{p}(p, p, p - 1)$ stands for ``prime''.

As an array of CPMs, $\bH_{p}$ consists of $p$ row-blocks of CPMs 
and $p$ column-block of CPMs.  Two different rows in a row-block 
of $\bH_{p}$ have no position where they both have 1-entries. Hence, 
the $p$ rows in each row-block correspond to $p$ parallel lines 
in the QC-PaG$_{p}(p, p, p - 1)$. Therefore, the QC-PaG$_{p}(p, p, p - 1)$ consists 
of $p$ parallel bundles, each consisting of $p$ parallel lines.  
Let $\bM_0 , \bM_1 , \dots  , \bM_{p-1}$  denote the $p$ parallel bundles of lines 
in the QC-PaG$_{p}(p, p, p- 1)$.  For $0 \leq  i < p$, the $p$ lines in 
the $i$-th parallel bundle $\bM_i$ correspond to the $p$ rows in 
the $i$-th row-block of $\bH_{p}$. The $p$ lines in each parallel bundle 
contain all the $p^2$ points in the QC-PaG$_{p}(p, p, p -1)$. This partial geometry is a net.
 
The $p$ rows in $\bH_{p}$ that have 1-entries at the $j$-th position are 
in $p$ different row-blocks. These rows correspond to $p$ lines 
in an intersecting bundle of lines that intersect at the 
point $v_j$ corresponding to the $j$-th column of $\bH_{p}$. The $p$ lines 
in an intersecting bundle are in $p$ different parallel bundles of the QC-PaG$_{p}(p, p, p - 1)$.
 
The $p^2$ points in the QC-PaG$_{p}(p, p, p - 1$) can be divided into $p$ 
sets, $\bN_0, \bN_1, \dots  , \bN_{p-1}$, each consisting of $p$ points. 
For $0 \leq j < p$, the $p$ points in the $j$-th set $\bN_j$ correspond to 
the $p$ columns of the $j$-th column-block of $\bH_{p}$.  The $p$ points of a 
line $L$ are distributed in $p$ different sets, one and only one 
point in each set.  A point $v$ in the set $\bN_j$ that is not on a 
line $L$ is adjacent to one and only one point on $L$ in each set other than $\bN_j$.
 
It follows from the results on the structural properties given in 
\cite{joh} that the Tanner graph, denoted by $\mathcal{G}_{p,\text{PaG}}$, of the QC-PaG$_{p}(p, p, p -1)$ has girth 6 and contains 
$p^3 (p - 1)^2 (p - 2)/6$ cycles of length 6.

The protograph, denoted by $\mathcal{G}_{p,\text{proto,PaG}}$  of the QC-PaG$_{p}(p, p, p -1)$  can be constructed directly from the base matrix $\bB_{p}$ given by (\ref{eq:base}).
 It consists of $p$ VNs and $p$ CNs, labeled by $\Phi_0,  \Phi_1,  \ldots, \Phi_{p-1}$ and $\Omega_0, \Omega_1, \ldots , \Omega_{p-1}$, respectively.
 The $p$ VNs and $p$  CNs of the protograph $\mathcal{G}_{p,\text{proto,PaG}}$ correspond to the $p$ columns and $p$ rows of the base matrix $\bB_p$ of the QC-PaG$_{p}(p, p, p - 1$). 
Every VN is connected to  every CN. The edge $(j, i)$ connecting the VN $\Phi_j$ to the CN $\Omega_i$  is labeled by the integer $ij$ modulo $p$.

If we remove the first column and the first row from the base matrix $\bB_p$, we obtain a Latin square of order $p - 1$. The use of Latin squares for constructing LDPC codes was considered in \cite{lae, zha}.

\begin{exmp} \label{eg3}
Let $p = 3$ and let $GF(3)$ be the field to construct 
a QC-PaG$_{p}(3, 3, 2)$.  Using  (\ref{eq:base}), we find the 
base matrix for constructing the line-point adjacency matrix $\bH_{p}$ 
of the QC-PaG$_{p}(3, 3, 2)$ is
 \begin{equation}
\bB_{p} = \;
\begin{bmatrix}
0  &  0 &   0\\
0 & 1 & 2 \\
0 & 2 & 1
\end{bmatrix}.
\end{equation}

The $3 \times 3$ CPM-dispersion $\bH_{p}$ of $\bB_{p}$ is
\begin{equation}
\bH_{p} = \;
\begin{bmatrix}
1 & 0 & 0 &   1 & 0 & 0 &   1 & 0 & 0 \\
0 & 1 & 0 &   0 & 1 & 0 &   0 & 1 & 0 \\
0 & 0 & 1 &   0 & 0 & 1 &   0 & 0 & 1 \\
1 & 0 & 0 &   0 & 1 & 0 &   0 & 0 & 1 \\
0 & 1 & 0 &   0 & 0 & 1 &   1 & 0 & 0 \\        
0 & 0 & 1 &   1 & 0 & 0 &   0 & 1 & 0 \\
1 & 0 & 0 &   0 & 0 & 1 &   0 & 1 & 0 \\ 
0 & 1 & 0 &   1 & 0 & 0 &   0 & 0 & 1 \\
0 & 0 & 1 &   0 & 1 & 0 &   1 & 0 & 0 
\end{bmatrix},
\end{equation}
\vspace{.1in}

\noindent
which is exactly the line-point adjacency matrix of the PaG$(3, 3, 2)$
given by (\ref{eq:mat1}) in Example \ref{eg2}. The Tanner graph and the 
protograph of this partial geometry are shown in Fig. \ref{fig:tann} and Fig. \ref{fig:proto2}, respectively.
\end{exmp}

\subsection{Subgeometries}
 
Let $\tau$ be a  positive integer less than $p-1$. Suppose we delete $\tau$
column-blocks from $\bH_{p}$.  We obtain a $p \times (p - \tau )$  subarray, 
denoted by $\bH_{p} (p, p - \tau)$ , of $\bH_{p}$. It follows from the structural 
property developed above that $\bH_{p}(p, p - \tau)$ is the line-point 
adjacency matrix of a QC-PaG$_{p}(p, p - \tau , p - \tau -1)$
which has $p(p - \tau)$ points and $p^2$ lines, each point on $p$ lines 
and each line consisting of $p - \tau$ points. The connection number 
of a point in the QC-PaG$_{p}(p, p- \tau, p - \tau -1)$  
is $p - \tau -1$. The QC-PaG$_{p}(p, p - \tau , p - \tau -1)$
is a {\it subgeometry} of the QC-PaG$_{p}(p, p, p -  1)$. Its Tanner graph (or protograph) is 
a subgraph of the Tanner graph (or protograph) of the QC-PaG$_{p}(p, p, p - 1)$.
 
Therefore, for a given prime field, we can construct a family of QC-PaGs.

\subsection{ QC-LDPC Codes on the QC-PaG$_{p}(p, p, p - 1)$ }
\label{sec:qcpag}
 
For $1\leq k, r \leq p$, let $\bB_{p} (k, r)$ be a $k \times r$ 
submatrix of the matrix $\bB_{p}$ of (\ref{eq:base}).  CPM-dispersing 
the entries of $\bB_{p} (k, r)$, we obtain a $k \times r$  array $\bH_{p} (k, r)$ of CPMs of 
size $p \times p$ which is a subarray of $\bH_{p}$. The array 
$\bH_{p} (k, r)$ is a $kp \times rp$ matrix 
with column and row weights, $k$ and $r$, respectively. The null space 
of $\bH_{p} (k, r)$ gives a $(k, r)$-biregular QC-LDPC code, denoted by $C_{p,qc} (k, r)$, 
of length $rp$ with rate at least $(r - k)/r$.  Therefore, for a given
prime field, a family of QC-LDPC codes of various lengths and rates can 
be constructed. The Tanner graph of $C_{p,qc} (k, r)$ has  girth 
of at least 6.
 
Express $\bB_{p} (k, r)$ as $\bB_{p}(k, r) = [b_{i,j} ]_{0 \leq i < k , 0 \leq j < r}$. 
Suppose an entry in $\bB_{p}(k, r)$  is replaced by the zero element of GF$(p)$. 
In the CPM-dispersion $\bH_{p}(k, r)$  of $\bB_{p}(k, r)$, this replacement 
results in replacing a $p \times p $ CPM in $\bH_{p}(k, r)$  by 
a $p \times p $ zero matrix (ZM) which is referred to as 
masking \cite{diao,qiuj,tan,xu}. Let $\lambda$  be a nonnegative integer less 
than the number of total nonzero entries in $\bB_{p} (k, r)$. The replacement 
of $\lambda$ nonzero entries in $\bB_{p}(k, r)$ by $\lambda$ zeros amounts to 
replacing $\lambda$ CPMs 
by $\lambda$ ZMs at the locations in $\bH _{p}(k, r)$ corresponding to the locations 
of the $\lambda$ entries in $\bB_{p}(k, r)$  which are replaced by zeros. 
Masking $\lambda$ CPMs in $\bH_{p}(k, r)$ amounts to removing $\lambda p $ edges 
from the Tanner graph $\cG_{p}(k, r)$ associated with $\bH_{p} (k, r)$. Removing 
these edges in $\cG_{p}(k, r)$ may break many short cycles in $\cG_{p}(k, r)$. 
As a result, the resultant Tanner graph $\cG_{p,\text{mask}} (k,r)$ may have a 
much smaller number of short cycles, or a larger girth, 
or both. The subscript ``mask'' stands for ``masking''. In choosing 
the entries in $\bB_{p}(k, r)$ to be masked, we have to avoid 
disconnecting the Tanner graph of $\bH_{p} (k, r)$.
 
The operation of masking $\bB_{p} (k, r) = [b_{i,j}]_{0 \leq i < k, 0 \leq j < r}$ 
can be modeled mathematically. Let 
$\bZ (k, r) = [z_{i,j}]_{0 \leq i < k , 0 \leq j < r}$
be a $k \times r$ matrix with the zero element and unit element of GF$(p) $
as entries. Define the following product of $\bZ (k, r)$ 
and $\bB_{p}(k, r): \bB_{p,\text{mask}}(k, r) = \bZ (k, r) \odot  \bB_{p} (k, r) 
= [z_{i,j} b_{i,j} ]_{0 \leq i < k , 0 \leq j < r}$ (the\emph{ Hadamard product})
where $z_{i,j} b_{i,j} = b_{i,j}$  if $z_{i,j} = 1$ and $z_{i,j} b_{i,j} = 0$ 
if $z_{i,j} = 0$. 
In this matrix product operation, entries in $\bB_{p}(k, r)$  at the 
locations corresponding to the locations of zero-entries 
in $\bZ (k, r)$  are replaced (or masked) by $0$'s. The CPM-dispersion 
of $\bB_{p,\text{mask}}(k, r)$  gives a $k \times r$ masked array $\bH_{p,\text{mask}} (k, r)$ of CPMs 
and ZMs of size $p \times p$. We call $\bZ (k, r)$ and $\bB_{p,\text{mask}}(k, r)$ the 
{\it masking matrix} and the {\it masked base matrix}, respectively. The null 
space of $\bH_{p,\text{mask}}(k, r)$ also gives a QC-LDPC code, 
denoted by $C_{p,qc,\text{mask}} (k, r)$.

\begin{exmp} \label{eg4}
Consider the $127 \times 127$ base matrix $\bB_{p}$ over GF$(127)$ given 
in (\ref{eq:base}). It contains the 
following $4 \times 8$ submatrix $\bB_{p}(4,8)$ (rows and columns 
chosen at random):
 \[
\bB_{p} (4,8) = \;
\begin{bmatrix}
2 &   83 &  33 &   46  &  36  &  94 &   42 &  86 \\
109 &  15 &   84  &   94 &   57  &  43 &  3 &  115 \\
112 &  76  &  70 &   36 &  111 &  57 &   66 &   117 \\
31 &  80 &   67 &   78 &    50 &  60 &  16 &  63 
\end{bmatrix}.
\]
 
We design the following masking matrix:
 \[
\bZ (4,8) = \;
\begin{bmatrix}
1 & 0 & 1 & 0 & 1 & 1 & 1 & 1 \\
0 & 1 & 0 & 1 & 1 & 1 & 1 & 1 \\
1 & 1 & 1 & 1 & 1 & 0 & 1 & 0 \\
1 & 1 & 1 & 1 & 0 & 1 & 0 & 1 
\end{bmatrix}.
\]
 
Masking $\bB_{p} (4,8)$ with $\bZ (4,8)$  gives the matrix $\bB_{p,\text{mask}}(4,8)$. 
Replacing each nonzero entry in $\bB_{p,\text{mask}} (4,8)$  by its 
corresponding $127 \times 127 $
CPM and each 0-entry by a $127 \times 127$ ZM, we obtain a masked $4 \times 8$ 
array $\bH_{p,\text{mask}} (4,8)$.  It is a $508 \times 1016$ matrix with column and row 
weights 3 and 6, respectively. The null space of $\bH_{p,mask} (4,8)$ 
gives a $(3,6)$-biregular binary $(1016, 508)$ QC-LDPC code with rate 0.5.  The                                                                                                       
Tanner graph of this code has girth 8 and contains 889 cycles 
of length 8.  The bit and block error performances of the code 
decoded with 5, 10 and 50 iterations of the min-sum algorithm (MSA) \cite{che} are shown in Fig. \ref{fig:graph3}.
We see that the code has a very low error-floor.  Included in the same 
figure are the bit and block performances of an LDPC code constructed 
using the progressive edge growth (PEG) algorithm \cite{hu}.  The PEG code is only decoded 
with 50 iterations of the MSA.  We see that the code constructed 
based on GF$(127)$ outperforms the PEG code below the bit error rate (BER) of $10^{-7}$.

\begin{figure}[!h]
\centering
\includegraphics[width=3.5in]{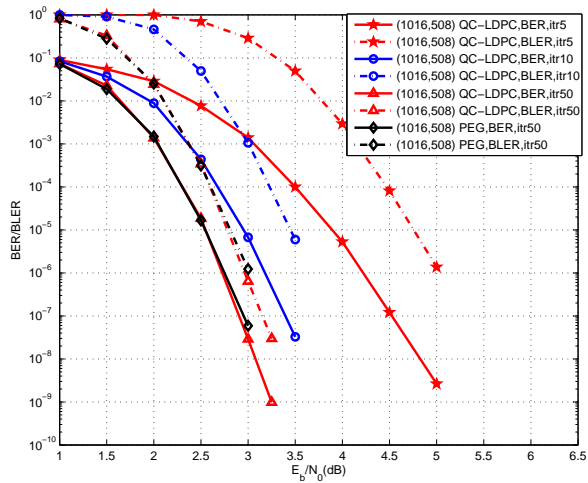}
\caption{Performance curves for the $(1016,508)$ QC-LDPC code and the PEG code given in Example \ref{eg4}.}
\label{fig:graph3}
\end{figure}

\end{exmp}

As noted, a special case of the partial geometries is when $\delta =1$, the generalized
quadrangles (GQs). In this case there are no triangles in the geometry 
and hence the associated graph has girth 8. 
There have been numerous studies on the use of GQs in coding theory
(although none considered their trapping sets to our knowledge).
The works include \cite{bag,bag1,bag2,liu,liu1,sas,sin,pep,von,von1}.
Some comments on the codes from GQs derived from these papers are given.
The paper of Liu et al \cite{liu1} contains a very useful summary of
the minimum distance bounds and rates of the various codes resulting from
using GQs for code constructions. Among other things, that work 
established the following two important results:

\begin{enumerate}
\item A code whose parity-check matrix is the adjacency matrix of a
generalized $d$-gon, for $n$ even, has only codewords of even weight.
\item A code whose parity check-matrix is the adjacency matrix of a
GQ is quasi-cyclic.
\end{enumerate}

This last point is important for code implementation. Its proof considers
the transitivity of the automorphism group of the quadrangle which
produces different sized circulant blocks for the different constructions.
The reader is referred to \cite{liu1} for details.

\section{A Class of  QC-PaGs Constructed Based On Cyclic Subgroups of Finite Fields} \label{sec:cyclegroup}
\vspace{.2in}

 In this section, we present another class of QC-PaGs which are constructed based on the cyclic subgroups of finite fields.

\subsection{ Construction}

Let GF($q$) be finite field with $q$ elements where $q$ is a power of a prime. Let $t$ be a prime factor of $q - 1$ and $q - 1 = ct$. If $\alpha$ is a primitive element of GF($q$), then $\beta=\alpha^{c}$ is an element of order $t$ in GF($q$). The set $\bS  = \{1, \beta, \beta^{2}, \ldots, \beta^{(t-1)} \}$ forms a cyclic subgroup of GF($q$) of order $t$.

 Form the following $t \times t$ matrix:
\begin{equation}
\label{eq:base_rs}
\bB_{c} = \;
\begin{bmatrix}
1  &  1 &  1 & \cdots & 1\\
1  &  \beta &  \beta^{2} & \cdots & \beta^{t-1}\\
1  &  \beta^{2} &  (\beta^{2})^{2} & \cdots & (\beta^2)^{t-1}\\
\vdots & \vdots & \vdots & \cdots & \vdots \\
1  &  \beta^{t-1} &  (\beta^{t-1})^{2} & \cdots & (\beta^{t-1})^{t-1}\\
\end{bmatrix}.
\end{equation}

Express the above matrix in the form $\bB_{c} = [\beta^{ij}]_{0 \le i, j < t}$ whose $(i, j)$-th element is $\beta^{ij}$.  The subscript ``$c$" stands for ``cyclic group". All the entries of $\bB_{c}$ are elements of the cyclic subgroup $\bS$. Let $\bP$ denote the CPM of size $t \times t$ whose first (or the top) row is the $t$-tuple $(0, 1, 0, \cdots , 0)$ with the single 1-component at the position-1 (the positions are labeled from 0 to $t  - 1$). For $0 \le \ell < t$, the $\ell$-th power of $\bP$, denoted by $\bP^{\ell}$ is a CPM of size $t \times t$ whose top row has its single 1-component at the position-$\ell$. Then, $\{\bI, \bP, \ldots, \bP^{(t - 1)}\}$ is a cyclic matrix group of order $t$ with $\bP$ as a generator and $\bP^t =\bP^0 =\bI$.

For $0 \le \ell < t$, we represent the element $\beta^{\ell}$ in the cyclic group $\bS$ of GF($q$) by the CPM $\bP^{\ell}$. This representation is one-to-one and $\bP^{\ell}$ is the CPM-dispersion of $\beta^{\ell}$.

Let $\bH_{c} = \bB_{c}(\bP)$ denote the CPM-dispersion of the matrix $\bB_{c}$, i.e., for $0 \le i, j < t$, the entry $\beta^{ij}$ of $\bB_{c}$ at the location $(i, j)$ is dispersed into the CPM $\bP^l$ where $\ell = (ij)~\text{modulo}~t$. Then, $\bH_{c} = \bB_{c}(\bP)$ is a $t \times t$ array of CPMs of size $t \times t$.  This array $\bH_{c} = \bB_{c}(\bP)$ is a $t^2 \times t^2$ binary matrix which is the line-point adjacency matrix of the QC-PaG$_{c}(t, t, t - 1)$.  The result can be established using either Theorem \ref{thm:one} or Theorem \ref{thm:two}, but Theorem \ref{thm:two} is used for illustrative purposes. Notice that $\bP^{T} = \bP^{-1}$. Consider the product of  $\bB_{c}(\bP)$ and its transpose $[\bB_{c}(\bP)]^{T}$. Express the product $\bB_c(\bP)[\bB_{c}(\bP)]^{T}$ as a $t \times t$ array of $t \times t$ matrices. Then, we have
\begin{equation*}
\bA = [\bA_{r,s}]_{0 \le r, s<t} = \bB_{c}(\bP)[\bB_{c}(\bP)]^{T}. 
\end{equation*}

For $0 \le r, s < t$, the constituent matrix $\bA_{r,s}$ of $\bA$ is
\[
\begin{array}{rl}
\bA_{r,s} & = \sum_{u=0}^{t-1} \bP^{uj} \left( \bP^T \right)^{us}  \\
        & = \sum_{u=0}^{t-1} \bP^{u(r-s)} = 
\left\{
\begin{array}{rl}
\bJ & \text{if} ~~ r \neq s \\
\rho \bI & \text{if} ~~ r=s
\end{array}
\right.
\end{array}
\]
where $\bJ$ is the $t \times t$ all ones matrix and $\bI$ the $t \times t$ identity matrix. Suppose the $i$-th column-block of CPMs of $\bB(\bP)$ is multiplied by $\bI + \bP^{j}$ to give $\bA^{(i,j)} = (\bA'_{r,s})$ where
\begin{equation*}
\bA'_{r,s} = \bA_{r,s} + \bP^{i(r-s)+j}. 
\end{equation*}

Thus, each CPM of each row-block of $\bB_{c}(\bP)$ has a CPM added to it. For a non-diagonal circulant ($r \neq s$), this means each row has $(t - 1)$ 2's with all remaining elements in the row being 0 or 1. It follows from Theorem \ref{thm:two} that the array $\bH_{c} =\bB_{c}(\bP)$, as a $t^2 \times t^2$ matrix, is the line-point adjacency matrix of a partial geometry PaG$_{c}(t, t, t - 1)$. The PaG$_{c}(t, t, t - 1)$ is a QC-PaG consisting of $n = t^2$ points and $m = t^2$ lines with connection number $\delta = t - 1$. This partial geometry PaG$_{c}(t, t, t - 1)$ is also a net.

If we want to use Theorem \ref{thm:one} to prove $\bH_c = \bB_c(\bP)$ is the line-point adjacency matrix of the partial geometry PaG$_{c}(t, t, t - 1)$, we need to show that any $2 \times 2$ submatrix of $\bB_c$ is nonsigular.  This can be done easily. It follows from \cite[Corollary 1]{QiuQin} that $\bH_c$ as a $t^2 \times t^2$ matrix satisfies the RC-constraint. Consequently, it follows from Theorem \ref{thm:one} that $\bH_{c} =\bB_{c}(\bP)$ is the line-point adjacency matrix of the partial geometry PaG$_{c}(t, t, t - 1)$.

The above construction gives another class of QC-PaGs whose line-point adjacency matrices are arrays of CPMs.  For $0 \le \tau < t$, if we delete $\tau$ column-blocks from $\bH_{c} =\bB_{c}(\bP)$, we obtain a $t \times (t - \tau)$ array $\bH_{c}(t, t - \tau)$ of CPMs of size $t \times t$ which is a $t^2 \times t(t - \tau)$ matrix. $\bH_{c}(t, t - \tau)$ is the line-point adjacency matrix of a partial geometry PaG$_{c}(t, t - \tau, t - \tau - 1)$ which is a subgeometry of the PaG$_{c}(t, t, t - 1)$.

Notice that all the entries in any row or column, other than the 0-th row and column, are different, since the order of $\beta$ is a prime. By removing the 0-th row and column from $\bB_{c}$, we obtain a Latin square of order $t-1$.

\subsection{QC-LDPC Codes on the QC-PaG$_{c}(t, t, t-1)$}

For $1 \le k, r \le t$, let $\bB_{c}(k, r)$ be a $k \times r$ submatrix of the matrix $\bB_c$ of (9). CPM-dispersing the entries of $\bB_c(k, r)$, we obtain a $k \times r$ array $\bH_c(k, r)$ of CPMs of size $t \times t$ which is a subarray of $\bH_c$. The array $\bH_c(k, r)$ is a $kt \times rt$ matrix with column and row weights, $k$ and $r$, respectively. The null space of $\bH_c(k, r)$  gives a $(k, r)$-biregular QC-LDPC code, denoted by $\mathcal{C}_{c,qc}(k, r)$, of length $rt$ with rate at least $(r - k)/r$. The Tanner graph of $\mathcal{C}_{c,qc}(k, r)$ has girth of at least 6.

In the following, we give an example to illustrate the construction of a QC-LDPC code with QC-structure based on the QC-PaG$_{c}(t, t, t-1)$.

\begin{exmp} \label{eg5}
Consider the field GF($2^7$). Since $2^7 - 1 = 127$ is a prime, it cannot be factored. So, we choose $t = 127$. Let $\alpha$ be a primitive element of GF($2^7$). Then, the set $\bS = \{1, \alpha, \alpha^2, \ldots, \alpha^{126}\}$ forms the only cyclic subgroup of GF($127$). The order of $\bS$ is 127. With the choice of $t = 127$, we construct a $127 \times 127$ base matrix $\bB_c =  [\alpha^{ij}]_{0 \le i, j < 127}$ in the form of (\ref{eq:base_rs}). All the 127 nonzero elements in GF($2^8$) appear in each row and each column of $\bB_c$.  Dispersing each entry in $\bB_c$ into a $127 \times 127$ CPM, we obtain a $127 \times 127$ array $\bH_c = \bB_c(\bP)$ of CPMs of size $127 \times 127$.  The array $\bH_c$ is a $16129 \times 16129$ matrix with both column and row weights 127.  This matrix is the line-point adjacency matrix of a QC-PaG$_{c}(127, 127, 126)$ with 16,129 points and 16,129 lines with connection number 126.
 
For $1 \le k, r \le t$, let $\bB_{c}(k, r)$ be a $k \times r$ submatrix of the base matrix $\bB_c$. Then, the null space of the CPM-dispersion of $\bB_{c}(k, r)$  gives a QC-LDPC code of length $127r$ whose Tanner graph has girth of at least 6.
 
Label the rows of the base matrix $\bB_c$ from $0$ to $126$. Suppose we take row-1 to row-6 from $\bB_c$ to form the following $6 \times 127$ submatrix $\bB_c(6,127)$ of $\bB_c$:

\begin{equation}
\label{eq:base_16129}
\bB_{c}(6,127) = \;
\begin{bmatrix}
1  &  \alpha &  \alpha^2 & \cdots & \alpha^{126}\\
1  &  \alpha^{2} &  (\alpha^{2})^{2} & \cdots & (\alpha^2)^{126}\\
1  &  \alpha^{3} &  (\alpha^{3})^{2} & \cdots & (\alpha^3)^{126}\\
1  &  \alpha^{4} &  (\alpha^{4})^{2} & \cdots & (\alpha^4)^{126}\\
1  &  \alpha^{5} &  (\alpha^{5})^{2} & \cdots & (\alpha^5)^{126}\\
1  &  \alpha^{6} &  (\alpha^{6})^{2} & \cdots & (\alpha^6)^{126}\\
\end{bmatrix}.
\end{equation}

Notice that matrix $\bB_{c}(6,127)$ given by (\ref{eq:base_16129}) is actually the parity-check matrix of the $(127, 121, 7)$ Reed-Solomon code of symbol length 127 over GF($2^7$) with minimum distance 7.  Dispersing each entry of  $\bB_{c}(6,127)$ into a $127 \times 127$ CPM, we obtain a $6 \times 127$ array $\bH_{c}(6,127)$ of CPMs of size $127 \times 127$ which is a subarray of $\bH_c$. It is a $762 \times 16129$ matrix with column and row weights 6 and 127, respectively.  The rank of this matrix is $757$. Then, the null space of $\bH_c(6,127)$ gives a $(6,127)$-biregular $(16129, 15372)$ QC-LDPC code $\mathcal{C}_{c,qc}(6,127)$ with rate $0.953$.

The bit and block error performances of this code decoded with the MSA are shown in Fig. \ref{fig:per16129} (computed with an FPGA decoder). We see that the code achieves a BER of $10^{-15}$ and a block error rate (BLER) of almost $10^{-12}$ without a visible error-floor. It has a beautiful waterfall performance. Fig. \ref{fig:per16129} shows the error performances of the code decoded with 5, 10 and 50 iterations of the MSA. We see that the decoding of the code converges very fast. At the BER of $10^{-12}$, the performance gap between 5 and 50 iterations of the MSA is about $0.5$ dB. The performance curves of the code decoded with 10 and 50 iterations of the MSA almost overlap all the way down to the BER of $10^{-12}$.

\begin{figure}[!h]
\centering
\includegraphics[width=3.5in]{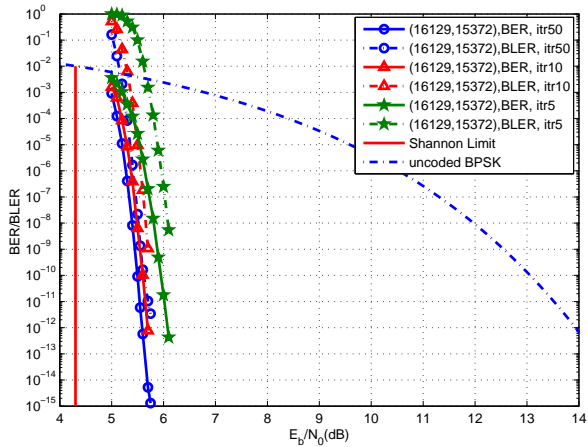}
\caption{Performance curves for the (16129,15372) QC-LDPC code given in Example \ref{eg5}.}
\label{fig:per16129}
\end{figure}

\end{exmp}
\vspace{.2in}

\section{Trapping Sets and Finite Gepmetries} \label{sec:trap}
\vspace{.2in}

Let $\cG$ be the Tanner graph of a binary LDPC code where the code is defined as the null
space of an $m \times n$ binary matrix. It is assumed that the $n$ vertices on
the left of $\cG$ are VNs of the code and the
$m$ vertices on the right are CNs. The code is to be decoded
via a BP decoder, such as the {\it sum-product algorithm}
(SPA) or the MSA. The behavior of these algorithms
depends on the cycle structure of the code as the iterations of the
algorithm attempt to converge to a codeword.

A trapping set for the code is defined as follows \cite{lae,ric}:
\begin{defn}
\begin{enumerate}
\item A $(\kappa , \tau )$ trapping set for the code defined
by the Tanner graph $\cG$ is a subset $\Delta, \mid \Delta \mid = \kappa$, of VNs such
that the subgraph of $\cG$ induced by the set $\Delta$,
denoted by $\cGD$, has exactly $\tau$ odd degree CNs (and
an arbitrary number of even degree CNs).
\item The trapping set is said to be {\it elementary} if all the associated
CNs in $\cGD$ have degree 1 or 2.
\item The trapping set is called {\it small} if $\kappa \leq \sqrt{n}$
($n$ is the code length) and $\tau / \kappa \leq 4$.
\end{enumerate}
\end{defn}

A good description of the motivation for these definitions is given
in \cite{lae,ric} and the reader is referred there
for the discussion. In essence, if there is a small trapping set,
there is a relatively higher probability that the BP algorithm will
fail to converge as it iterates. Beyond trapping sets, the notion
of an {\it absorbing set} was introduced in \cite{dol} to aid
in the convergence analysis but as noted, this notion will not be
considered in this work.

Trapping sets have been discussed for a variety of LDPC codes
obtained from combinatorial structures (in a variety of ways)
(see the works \cite{diao,qiuj,lin,kim,lae,lae2,lae3}).
The question of interest is to discuss them in light of the structure of
the finite geometric structure of the code to determine if sharper
bounds can be found over those for other codes.

A useful general result is given in Theorem 2 of \cite{diao,qiuj}. It shows
that if $\cG$ is the Tanner graph of an LDPC code with girth of at least 6,
with VNs of degree $\gamma$, and $\cG$ contains a
$(\kappa , \tau )$ trapping set and $\kappa < \gamma$, then
\[
\tau \geq (\gamma +1 - \kappa ) \kappa.
\]

A partial geometry code is the dual space to the row space of
the $m \times n$ adjacency matrix of the partial geometry PaG$(\gamma,
\rho, \delta)$.

The trapping sets of such codes were examined in
\cite{diao,qiuj} and a brief discussion of those results was given.  It
is noted that the geometric structure of the partial geometry allows
a better analysis of the trapping sets over matrices derived from
less structured objects. Let $\Delta$ be a subset
of VNs of size $\kappa$ and $\cGD$ the subgraph of the Tanner
graph of a code generated by a partial geometry. The properties of the
trapping set depends on the structure of this subgraph induced
by $\Delta$. Since the CNs generated by $\Delta$
correspond to lines through the points of $\Delta$, it will generate
a $(\kappa , \tau)$ trapping set if there are precisely
$\tau$ lines in $\cGD$ passing through an odd number of points of $\Delta$.
Let $m_i$ be the number of lines of $\cGD$ passing through 
$i$ points of $\Delta$ (which is the number of CNs of
degree $i$). Then:

\begin{thm} \label{thm:three}(\cite[Theorem 3]{qiuj})
Let $\cG_{\text{PaG}}$ be the Tanner graph of a PaG$( \gamma , \rho , \delta )$.
If $\Delta$ is a $(\kappa , \tau )$ trapping set and $\kappa \leq \gamma$
then:
\[
\tau  \geq (\gamma +1 -\kappa) \kappa + \sum_{i~ \text{odd}} (i-1)^2 m_i + \sum_{i~\text{even}} i(i-2) m_i
\]
and equality holds if $\delta = \rho$ and the sums go to
$2 \lfloor (\kappa+1)/2 \rfloor -1$ and $2 \lfloor \kappa /2 \rfloor$, respectively.
\end{thm}

Note that the first term on the right hand side of this expression is
the general bound noted earlier. Also the number of 
edges in the graph generated by a given set of $\kappa$ VNs is,
by definition of the partial geometry, $\kappa  \gamma$. Hence
\[
 \sum_{\text{all} ~ i} i m_i = \kappa \gamma \; \; \text{and} \; \; \sum_{\text{odd} ~i} m_i = \tau .
\]

In the particular case of a net the bound can be improved somewhat.
Let $L_0 , L_1 , \dots , L_{\rho-1}$ be a set of parallel
lines in PaG$( \gamma , \rho , \gamma -1)$ and for a set $\Delta$
of VNs let $\kappa_i = \mid \Delta \cap L_i \mid$. Then:

\begin{thm} \label{thm:four}
If the set $\Delta$ of VNs of a net is a $(\kappa , \tau )$ trapping set, then
\[
\begin{array}{rl}
\tau & \geq (\gamma -1) \kappa - \kappa^2 \\
      & + \sum_{i=1}^{\rho} \kappa_i^2 + \mid \{ \ell :~ 1 \leq \ell \leq \rho ,~ \kappa_{\ell} ~ \text{odd} \} \mid.
\end{array}
\]
\end{thm}

It is noted that the bound agrees with the previous one
whenever $\kappa_i \leq 2$ for all $i$ and improves on it in other cases.
A lower bound on the average size of $\tau$ is also given in \cite[Corollary 1]{qiuj}.

It was observed \cite{lin} that the structure of the partial geometric
code allows comments on trapping sets for certain types of sets of
VNs and a few of these comments are noted here.

A {\it partial ovoid} of a partial geometry PaG$( \gamma , \rho , \delta )$ is a set $\bS$ of points (VNs) such that every line in the geometry is incident with at most one point of $\bS$.
Such a set can have at most $1+ (\gamma -1)( \rho -1)$ points.
Thus CNs (lines) corresponding to such a set of size $\kappa$ 
has $\kappa \gamma $ lines that intersect the
$\kappa$ VNs and each such CN has degree 1. 
Hence $m_1 = \kappa \gamma$ and all other $m_i's$ are 0. 
The above bounds are difficult to work with in this case but in the case that $t>3$
the ratio $\kappa / \tau = 1+t > 4$ and
such a set of VNs cannot be a small trapping set.

Consider a set of $\kappa$ VNs that are colinear, i.e., one line
contains all $\kappa$ points. The subgraph generated by such a configuration has 
$\tau = (\gamma -1) \kappa + 1$ CNs, $m_{\kappa} =1$, $m_1 = \kappa (\gamma -1)$
and $\tau / \kappa = (\gamma -1) + 1 / \kappa$.
For $\gamma > 4$ this does not correspond to a small trapping set. 

From these arguments it is clear that the smallest trapping sets
will arise from CN sets that contain large numbers of lines
between the $\tau$ points. An extreme case of this is a clique.
A {\it clique} of a partial geometry is a set of VNs which are mutually 
colinear. A clique of size $\kappa$ in the partial geometry
will generate a subgraph with $(\gamma  - \kappa) \kappa = m_1$ CNs of degree 
$1$ and $\binom{\kappa}{2} = m_2$ of degree $2$. 
The maximum size of a clique for a given partial geometry is not known, but, depending 
on the values of $\tau$, $\gamma$ and $\kappa$, they might form 
a significant trapping set since, from the above argument, 
assuming a clique of size $\kappa$ exists, $\tau / \kappa = \rho - \kappa$.
A clique of size $>2$ is not possible for GQs since it implies forbidden triangles.

The cases above are extreme but serve as demonstrations that the bounds on the
sizes of the trapping sets might be sharpened for the case of partial 
geometries and GQs.

\section{Expansion properties of the graphs of partial geometries} \label{sec:graphs}

The notion of expansion in coding theory originated
in the work of Tanner \cite{tann2} (under the name of 
the strongly related concept of {\it superconcentrators}). That work 
also discovered the important relationship between
the second largest eigenvalue of the adjacency graph of the code
and its expansion properties, a relationship that has been
widely exploited by mathematicians and computer science researchers
since that time. These issues are examined further here.

Codes that can be encoded and decoded in linear time can be 
constructed by means of codes with suitable expansion properties 
\cite{sips}, the {\it expander codes}, and such codes and their graphs have 
received considerable attention over the past two decades. The codes
were not LDPC codes and their decoding algorithm was not a 
message-passing one. However, it can be argued \cite{tann2} that expansion 
properties are also of interest in belief propagation
decoding algorithms for LDPC codes since the notion of graph expansion 
can be interpreted as a measure of connectivity and randomness 
of the graph, desirable properties for efficient decoding properties.
Tanner used such arguments in his seminal paper \cite{tann}. In \cite{tann1}
he showed that the minimum distance properties of graph-based codes
with small second eignevalue relative to the largest eigenvalue
were also good. However, it is in the work of Burshtein and Miller 
\cite{bur} that a more direct relationship between expander graphs and
the performance of message-passing decoding algorithms is established.
That work uses expander-based arguments to establish that for sufficiently
long block lengths, once a message-passing algorithm corrects
a sufficiently large fraction of errors, it will eventually correct
all errors. The argument considers Gallager hard and soft decoding algorithms
but will be applicable for a wider class as well.

Since the second eigenvalue property is important for codes from several 
points of view, it seems worthwhile to examine the expansion
properties of the graphs from partial geometries, a problem that
is addressed in this section, with the thought of using this property
as a possible distinguisher of codes for further examination.
The extent to which these arguments are viable would have to 
be confirmed with simulation.

The eigenvalues of graphs associated with partial geometries
are well known and the contribution of this section is to modify 
the known results to the purpose of interest, namely the expansion properties
of graphs from these geometries. A brief overview of the key ideas involved is given.

The majority of works on this
topic consider regular graphs, graphs with each vertex having the same degree.
Our interest is solely in the  biregular bipartite case, defined as
bipartite graphs with $n$ left vertices, each of constant degree $c$
and $m$ right vertices, each of degree $d$. When $m$ and $n$
are understood, we refer to this as a $(c,d)$-biregular bipartite graph. 
The recent work of \cite{hoh} is an important contribution to 
this problem as it is the first work known to the authors
that considers the expansion properties of these graphs, apart from the
original seminal work of Tanner \cite{tann2}. Large
classes of such graphs have been considered in the literature as 
Tanner graphs associated with partial geometries and other combinatorial
configurations. Much of the work on LDPC codes is in fact on such 
biregular graphs.

The notion of graph expansion is defined in several related ways. The following 
definition \cite{hoh} will be used  since other results of that work will be
of interest. The same notion was used in the original work of Tanner \cite{tann2},
which also considered biregular bipartite graphs.
Let a general graph $\cG = (\bV,\bE)$ (not necessarily biregular bipartite)
with set of vertices $\bV$ and edges $\bE$ be a connected
graph with no self loops. Define the {\it boundary} $\delta \bX$ 
of a subset $\bX \subset \bV$ with $\mid \bX \mid \leq \alpha \mid \bV \mid$ as the set of neighbors of $\bX$.
Define the {\it expansion coefficient} $c ( \alpha )$ 
for $\mid \bX \mid \leq \alpha \mid \bV \mid$, for some positive fraction
$\alpha$, of the graph by:
\[
c ( \alpha ) = \min_{\bm{\phi} \neq \bX \subset \bV} \frac{\mid \delta \bX \mid }{\min \{ \mid \bX \mid , \mid \bV \backslash \bX \mid \} },
\]
for set $\bX \subset \bV$. Interest is often in the case where $\alpha < 1/2$
where the minimization in the denominator is not necessary.

In the case that $\cG = (\bV,\bE)$ is a $k$-regular (each vertex is of degree $k$),
$\mid \bV \mid = n$), and let $\bA =[a_{ij}]_{0 \leq i,j < n}$ be its $n \times n$ adjacency matrix
(i.e., its point-point adjacency matrix where $a_{i,j} = 1$ if 
vertex $v_i$ is connected to vertex $v_j$). The matrix $\bA$ will have 
$n$ real eigenvalues which are listed in decreasing order
\[
\mu_0 \geq \mu_1 \geq \cdots \geq \mu_{n-1},
\]
and $n$ orthogonal eigenvectors.

If $\cG$ is $k$-regular then $\mu_0 = k$ and $\mu_{n-1} = - \mu_{0}$ 
if and only if it is bipartite. It can be shown that for any family 
of $k$-regular connected graphs with number of vertices tending to infinity
will have 
\[
\liminf_{n \Rightarrow \infty} \mu_1 \geq 2 \sqrt{k-1}.
\]
A finite connected $k$-regular graph will be called {\it Ramanujan}
if all of its eigenvalues other than $\pm k$ satisfy the bound
\[
\mu \leq 2 \sqrt{k-1}.
\]
Such a graph is described as having a ``small'' second eigenvalue.
The amount of literature on the search for graphs with this property
is very large.

As noted, the interest of much of the theory on LDPC codes
is focussed on the construction of $(c,d)$-biregular
bipartite graphs for use as Tanner graphs of the code and the
recent work \cite{hoh} is precisely on the expansion properties of such
graphs. This section considers that work in the light of 
the graphs from partial geometries, all of whose eigenvalues are known.

Let $\bV_1$ be the set of $c$-regular VNs and
$\bV_2$ the set of $d$ regular CNs.
Interest is in the case where the
expansion in the previous definitions is for sets of VNs and
$\bX \subset \bV_1$ and its boundary is a subset of the CNs $\bV_2$
and bounds on the resulting expansion coefficients is in terms of
the eigenvalues of the adjacency matrix.
This case was included in the interesting recent work \cite{hoh} 
and several results from that work will be of interest 
here. That work defined the $(c,d)$-biregular bipartite graph $\cG$ to be 
Ramanujan if 
\[
\mu_1 (\cG ) \leq \sqrt{c-1} + \sqrt{d-1},
\]
a natural extension of the regular graph case by setting $c=d$,
although not theoretically justified.

For coding applications, the parity check matrix $\bH$ is often the 
$m \times n$ matrix derived from a combinatorial configuration
or finite geometry, where the $m$ rows of $\bH$ are identified with the blocks of the
configuration or lines of the geometry and the columns of $\bH$ with the
points. It can be described as the line-point adjacency matrix of
the structure. The work of \cite{hoh} refers to this matrix as 
the {\it transfer matrix}.

In terms of this matrix $\bH$, the adjacency matrix of the configuration
or geometry which is $n \times n$, a point-point adjacency matrix,
(and not using the blocks as graph vertices) is easily seen to be 
\begin{equation}
\label{eq:srg}
\bA_1 = {\bH}^T \bH -  \gamma \bI,
\end{equation}
where there are $ \gamma$ lines intersecting a point and $\bI$ is a $n \times n$ identity matrix.

The related adjacency matrix $\bA$ which includes both the 
left vertices of the Tanner graph (VNs of the code) and the 
right vertices (CNs of the code), is of the form
\begin{equation}
\label{eq:tanner}
\bA = 
\begin{bmatrix}
{\bf O} & {\bH}^T \\
\bH & {\bf O}
\end{bmatrix}.
\end{equation}

This is the matrix and graph whose expansion properties are of interest.
Of course in the case of a $(c,d)$-biregular bipartite graph, the rows of
the $m \times n$ matrix $\bH$ have weight $d$ and columns have weight $c$.
Note that
\begin{equation}
\label{eq:pos}
{\bA}^2 =
\begin{bmatrix}
{\bH}^T \bH & {\bf O} \\
{\bf O} & \bH {\bH}^T
\end{bmatrix} .
\end{equation}

As discussed in \cite{joh,cam}, the eigenvalues of the adjacency matrix of
graphs of partial geometries are well known. Such combinatorial structures correspond
to two class association schemes and strongly regular graphs and the eigenvalues
of the adjacency matrix of the graphs of the PaG$( \gamma, \rho , \delta )$ 
$\bA_1$ of (\ref{eq:srg}) are given by:
\begin{equation}
\label{eq:srg1}
\begin{array}{rl}
(\rho -1) \gamma  & \text{multiplicity} \; 1, \\
\rho -1 - \delta  & \text{multiplicity} \; \frac{\gamma \rho (\gamma -1)(\rho -1)}{\delta  (\gamma + \rho -1 - \delta)}, \\
- \gamma  & \text{multiplicity} \; \frac{(\rho-1)  (\rho -\delta )((\gamma -1)(\rho -1)+\delta)}{\delta  (\gamma + \rho - 1 -\delta)} .
\end{array}
\end{equation}

It is clear \cite{joh} that the eigenvalues of $\bH^T \bH$ are those of the 
matrix $\bA_1$ of (\ref{eq:srg}), with $\gamma$ added to each eigenvalue, and with the 
same multiplicities,  i.e., the eigenvalues of $\bH^T \bH$ are $\gamma \rho$, $\gamma + \rho -1 - \delta$ and $0$.

It is noted in \cite{tann2} that if $\bx$ is an eigenvector of ${\bH}^T \bH$ with eigenvalue
$\lambda \neq 0$ then $\bx^{\prime} = \bH \bx$ is an eigenvector 
of ${\bH} {\bH}^T$ with the same eigenvalue. Thus ${\bH}^T \bH$ and ${\bH}{\bH}^T$ have the same eigenvalues
(with different multiplicities). It is not difficult to show that if the eigenvalues
of the matrix of interest $\bA$ in (\ref{eq:tanner}) are $\{ \mu_i , ~ i=0,1, \dots , n-1 \}$,
then the eigenvalues of ${\bH}^T \bH$ are $\{ \mu_i^2, ~ i=0,1, \dots , n-1 \}$ and the expansion properties of the
graph are determined by the ``second'' eigenvalue $\mu_1$. 

As $\bA$ is a real symmetric matrix we have the rank of $\bA$, $r_{\bA}$, is equal to the rank of ${\bA}^2$
and from (\ref{eq:tanner}), $r_{\bA} = 2r_{\bH}$. The eigenvalues of ${\bA}^2$ are those of ${\bH}^T \bH$ which are the same
(in magnitude, not necessarily in multiplicity) as those of ${\bH}{\bH}^T$. Thus, arguing in the reverse from the previous
paragraph, the eigenvalues of the matrix $\bA$ of (\ref{eq:tanner}) are $\pm \sqrt{\gamma \rho}$, $\pm \sqrt{\gamma + \rho -1 - \delta}$ and $0$.
Thus the second eigenvalue is $\sqrt{\gamma + \rho -1 -\delta}$.

The expansion coefficient of any $(c,d)$-biregular bipartite 
graph is given by \cite{hoh,tann2}
\begin{equation}
\label{eqn:exp}
c (\alpha ) \geq \frac{c^2}{\alpha cd + \mu_1^2 (1-\alpha )}
\end{equation}
and the importance of the magnitude of the second eigenvalue is seen, i.e.,
the smaller the size of $\mu_1$ the larger the expansion of the graph.

Such information is used in \cite{tann1} and \cite{joh} to determine bounds on the
minimum distance of the codes with parity-check matrix $\bH$. The interest here is
in the expansion properties of the code with parity-check matrix $\bH$.
Since the second eigenvalue of the matrix $\bA$ is $\sqrt{\gamma + \rho -1 -\delta}$,
this suggests that the graphs of partial geometries with $\delta$ as large
as possible, while still being partial geometries, will have the best expansion
properties, i.e., $\delta = \gamma -1$ corresponding to  nets.  Notice that the two classes of partial geometries constructed in this paper are two classes of nets. For the first class, $\gamma = \rho = p$ and $\delta = p-1$, and for the second class, $\gamma = \rho = t$ and $\delta = t-1$. Hence, they have good expansion properties. While the 
parameters of the partial geometry, $\gamma$, $\rho$  and $\delta$, are 
not independent, the conclusion is interesting and worthy of further consideration.

For interest, to conclude this section, the ratio of $\mu_1^2 / \mu_{\max}^2$ is compared
for biregular bipartite graphs and $k$-regular graphs. From the discussion,
for $k$-regular graphs, for the Ramanujan case, 
\[
\mu_1^2 /\mu_{\max}^2 =  2 (k-1) /k^2 \approx 2/k
\]
while for $(\gamma, \rho)$-biregular bipartite graphs from 
partial geometries the ratio is
\[
\mu_1^2 / \mu_{\max}^2 = \frac{\gamma + \rho - (\delta +1)}{\gamma \rho} = \frac{1}{\rho} + \frac{1}{\gamma} 
- \frac{(\delta +1)}{\gamma \rho}.
\]
If parameters can be chosen so that $\rho \approx \gamma \approx k$ and $\delta$ small,
the two ratios are similar. For larger values of $\delta$, however, such as for nets,
the ratio of the eigenvalues of the partial geometry is 
approximately $1/{\rho}$, suggesting graphs
with large $\rho$ and $\delta$ would have better expansion.

\section{Conclusion and Remarks}
 
Several aspects of codes derived from partial geometries have been considered.
New results on codes from partial geometries were given, including an interesting characterization of them in terms of arrays
of cyclic permutation matrices and two new and simple constructions. 
The trapping sets of codes from partial geometries were also investigated 
using the geometric 
properties of their constructions. Finally, comments were given on
the expansion properties of graphs from partial geometries.

\end{document}